\documentclass[smallextended]{svjour3}     % onecolumn (second format)

\smartqed  % flush right qed marks, e.g. at end of proof
\usepackage{graphicx}
\usepackage{amssymb}
\usepackage{xcolor}
\begin{document}

\title{Coherent caloritronics in Josephson-based nanocircuits %\thanks{Grants or other notes
}


\author{M. J. Mart\'inez-P\'erez         \and
        P. Solinas	\and
        F. Giazotto
}


\institute{M. J. Mart\'inez-P\'erez and F. Giazotto \at
              NEST, Instituto Nanoscienze-CNR and Scuola Normale Superiore, I-56127 Pisa, Italy \\
              \email{ mariajose.martinez@sns.it} 
              \email{ f.giazotto@sns.it}           %  \\
           \and
           P. Solinas \at
              SPIN-CNR, Via Dodecaneso 33, I-16146 Genova, Italy
}

\date{Received: date / Accepted: date}

\maketitle

\begin{abstract}
We describe here the first  experimental realization of a heat interferometer, thermal counterpart of the well-known superconducting quantum interference device (SQUID). These findings demonstrate, on the first place, the existence of  phase-dependent heat transport in Josephson-based superconducting circuits and, on the second place, open the way to novel ways of mastering heat at the nanoscale. Combining the use of external magnetic fields for phase biasing and different Josephson junction architectures we show here that a number of heat interference patterns can be obtained. The experimental realization of these architectures, besides being relevant from a fundamental physics point of view, might find important technological application as building blocks of phase-coherent quantum thermal circuits. In particular, the performance of two different heat rectifying devices is analyzed.
\end{abstract}

\section{Introduction}
\label{intro}

The Josephson effect has implications going beyond electrical transport \cite{Josephson}. In 1965, Maki and Griffin predicted that the interplay between Cooper pairs and quasiparticles in tunneling events would provide heat currents with quantum coherence \cite{Maki}. This phenomenon manifests itself as a phase-dependent component of the heat current flowing through a thermally-biased Josephson junction (JJ), which should, therefore, enable the realization of heat interference. Towards this end, a \textit{magnetic flux-controllable} superconducting heat interferometer was recently theoretically conceived \cite{GiazottoAPL12} and subsequently realized experimentally \cite{Giazottoarxiv,simmonds2012}. This achievement served to show, on the one hand, how quantum coherence %between two weak-linked superconducting condensates 
extends also to dissipative observables such as heat current. On the other, these devices might constitute the building blocks for the implementation of superconducting hybrid \emph{coherent caloritronic circuits} consisting of, for instance, thermal modulators, heat transistors, splitters, etc. 

The purpose of this paper is to review recent advances done within the emerging field of \emph{coherent caloritronics}. 
The paper is organized as follows: We dedicate Sec. \ref{sec00} to distinguish first between coherent caloritronics and thermoelectric effects also present in temperature-biased JJ. In Sec. \ref{sec02} we describe the general model used to derive the behavior of the \emph{heat} current flowing through the JJ. In Sec. \ref{sec1} a superconducting heat interferometer is analyzed theoretically and its operation is demonstrated experimentally. In Sec. \ref{sec2} we discuss the implementation of a double-loop heat modulator %thanks to which the flux-to-heat current transfer function can be tailored at will or even suppressed through the application of an extra control flux. 
An even more versatile device is theoretically analyzed in Sec. \ref{sec4}. It consist of a temperature-biased extended JJ that serves to produce coherent diffraction of the thermal currents. %The phase-dependent heat current in three specific junction geometries is further analyzed. 
In Sec. \ref{sec5} the remarkable heat rectification properties of NIS and SIS' junctions are addressed (where N stands for a normal metal, I represents a thin insulating layer and S and S' refer to two different superconducting electrodes). This review is enriched with realistic calculations showing the achieved  modulation of the electronic temperature of a normal metal electrode placed nearby each of the aforementioned structures. Finally, our results are summarized in Sec. \ref{Summary} along with a few final remarks.

\section{Preliminary considerations: thermoelectric effects in SNS junctions.}
\label{sec00}

It is convenient to begin by clearly stating the nature of the phenomena we will deal with along this manuscript. It is our goal to investigate the intriguing quantum behavior of the \emph{heat} current $\dot{Q}_{\rm tot}$ flowing through a temperature-biased JJ. Here, we will concentrate on JJs consisting of two superconducting electrodes separated by a thin insulating layer, i.e. tunnel SIS junctions. 

 These studies are not to be confused with thermoelectric effects also present in JJs \cite{aronov,Panaitov}. The latter concern the consequences of a thermal bias in the \emph{electric} current flowing through the junction and manifests as the generation of a potential difference of thermoelectric nature. Specifically, a thermovoltage $V_{\rm th}$ might arise across the junction when the electronic temperature difference between the superconducting electrodes exceeds a threshold value $\Delta T_{\rm th}$ \cite{schmidt}. These effects can be understood by assuming that the temperature drop across the junction gives rise to a normal electric current ($i^{\rm N}_{\rm th}$). As noted by Ginzburg \cite{Ginzburg44},  this current is shorted out in superconducting materials by the generation of a supercurrent ($i_{\rm th}^{\rm J}$) flowing in the opposite direction. As a result, the voltage difference generated across the junction remains zero. If, however, the temperature drop is large enough to generate $|i_{\rm th}^{\rm N}|=i_{\rm th}^{\rm J}>i_c$, $i_c$ being the critical current of the JJ, one enters into the dissipative regime leading to the appearance of $V_{\rm th}$. The smallness of this effect has complicated considerably its experimental observation, making it measurable only in certain kind of JJs such as the SNS type. In this way, Kartsovnik \textit{et al} observed the generation of $V_{\rm th} \sim 10^{-13}$ V upon temperature differences of the order of $\Delta T_{\rm th} \sim 10^{-5}$ K in Ta-Cu-Ta SNS sandwiches \cite{Kartsovnik}. Thermoelectric effects on response to an in-plane magnetic field were also studied by Ryazanov \textit{et al} in Ref. \cite{Ryazanov}. Specifically, $\Delta T_{\rm th}$ was found to depend on the modulus of the cardinal sine function of the magnetic flux piercing the junction leading to a Fraunhofer-like dependence. %These experimental results provided evidence of the existence of $V_{\rm th}$ upon the application of a sufficiently high temperature bias.

The influence of thermoelectric phenomena in SIS junctions was analyzed in Ref. \cite{Guttman0}. According to these calculations,  a temperature gradient of $\sim 10^{-1}$ K across a tunnel junction with normal state resistance $R_J = 100$ $\Omega$ will generate a Cooper pair thermocurrent $i_{\rm th}^{\rm J} \sim 10^{-9}$ A. The latter will imply, in turn, the existence of a tiny phase difference between the superconducting electrodes of only  $\Delta \varphi_{\rm th} \sim 10^{-4}$ rad that can be neglected for the considerations made here.

\section{Heat transport in JJs: equations.}
\label{sec02}

We will turn our attention now to the description of the main equations governing heat transport in JJs. 

\begin{figure}[t]
\centering
\includegraphics[width=.7\textwidth]{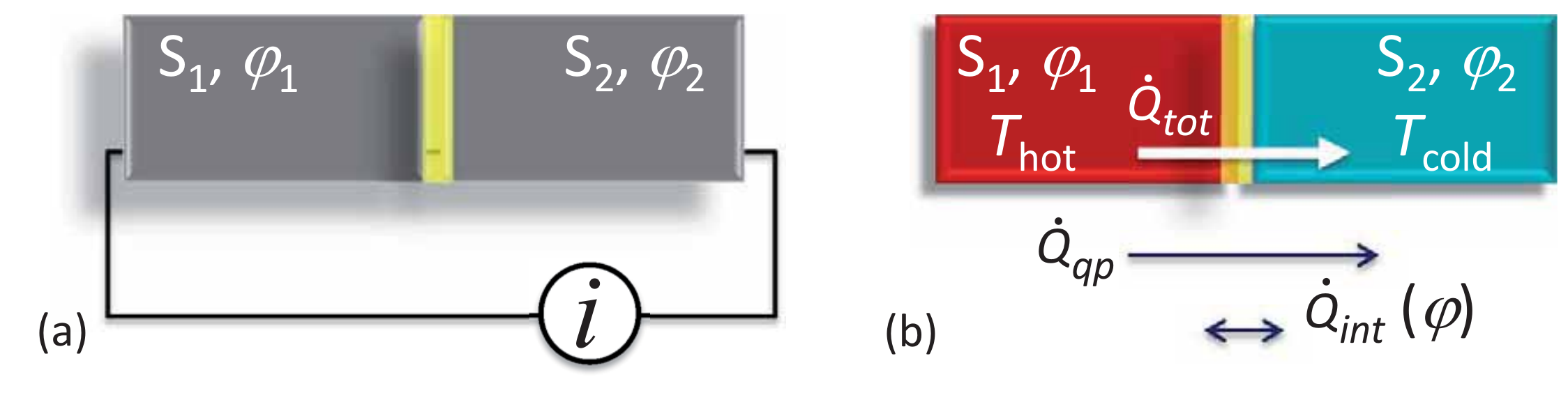}{\centering}
\caption{(color online) (a) Two superconductors S$_1$ and S$_2$ separated by a thin insulating layer constitute a conventional JJ. If a phase difference $\varphi =\varphi_1 - \varphi_2$ arises, an electric supercurrent $i_{\rm J} = i_c \sin \varphi$ will flow. (b) If S$_1$ and S$_2$ are kept at temperatures $T_{\rm hot}$ and $T_{\rm cold}$ (with $T_{\rm hot}\geq T_{\rm cold}$), respectively, a steady-state total heat current  $\dot{Q}_{tot}=\dot{Q}_{qp}-\dot{Q}_{int}\cos\varphi$ will flow from S$_1$ to S$_2$. %Here,  $\dot{Q}_{qp}$ is the usual heat flux carried by quasiparticles, $\dot{Q}_{int}$ is the phase-dependent part of the heat current peculiar of Josephson tunnel junctions.
}
\label{Fig01}
\end{figure}

We start by assuming a Josephson tunnel junction between two superconducting electrodes $S_1$ and $S_2$ characterized by a phase difference $\varphi = \varphi_1 - \varphi_2$ as schematized in Fig. \ref{Fig01}(a). Under such circumstances, an \emph{electric} supercurrent will flow that depends on $\sin \varphi$. Assume now that the JJ is electrically open but a thermal gradient is established by heating up the quasiparticles of $S_1$ up to $T_{\rm hot}$ while keeping quasiparticles in $S_2$ at temperature $T_{\rm cold}$. Then, a stationary heat current $\dot{Q}_{tot}$ will flow from the hot to the cold reservoir.

%____________________________________________________________________________________________________________________________________________________________

%____________________________________________________________________________________________________________________________________________________________

%____________________________________________________________________________________________________________________________________________________________

%____________________________________________________________________________________________________________________________________________________________

The explicit form of $\dot{Q}_{tot}$ was calculated by Maki and Griffin in 1965 \cite{Maki} giving.

\begin{equation}
\dot{Q}_{\rm tot}(T_{\rm hot},T_{\rm cold},\varphi)=\dot{Q}_{qp}(T_{\rm hot},T_{\rm cold})-\dot{Q}_{int}(T_{\rm hot},T_{\rm cold})\textrm{cos}\varphi.
\label{heatcurrent}
\end{equation}

 Equation (\ref{heatcurrent}) consists of two terms; $\dot{Q}_{qp}$ is the usual heat flux carried by quasiparticles \cite{RMP} and $\dot{Q}_{int}\textrm{cos}\varphi$ is the phase-dependent part of the heat current, which is peculiar of Josephson tunnel junctions.   Quoting the authors, the latter term ``gives rise to an oscillatory heat flux and is analogous to a similar term in conventional electric transport through Josephson junctions''. We stress the fact that, depending on $\varphi$,  the phase-dependent component of the heat current can flow backwards, i.e., opposite to the thermal gradient. The total heat current goes however from the hot to the cold reservoir, preserving the second principle of thermodynamics \cite{simmonds2012}. It is worthwhile to stress as well the absence of any Cooper pair-related energy term. This is so for, in a static situation, the superconducting condensate shall carry no entropy  \cite{Maki,Golubev2013}.

The quasiparticle term appearing in Eq. (\ref{heatcurrent}) reads \cite{Maki,Golubev2013,Guttman1,Guttman2,Sauls1,Sauls2}
\begin{eqnarray}
\dot{Q}_{qp}(T_{\rm hot},T_{\rm cold})=\frac{1}{e^2 R_J} \int^{\infty}_{0} d\varepsilon \varepsilon  \mathcal{N}_1 (\varepsilon,T_{\rm hot})  \mathcal{N}_2 (\varepsilon,T_{\rm cold}) \nonumber \\ \times  [f(\varepsilon,T_{\rm cold})-f(\varepsilon,T_{\rm hot})],
\label{qqp} \
\end{eqnarray}

where $\mathcal{N}_{k}(\varepsilon,T_k)=\frac{|\varepsilon|}{\sqrt{\varepsilon^2-\Delta_{k}(T_k)^2}} \Theta[\varepsilon^2-\Delta_{k}(T_k)^2]$ 
is the quasiparticle BCS normalized  density of states in S$_{_k}$ at temperature $T_k$ ($k=$hot, cold) and $\varepsilon$ is the energy measured from the condensate chemical potential. Furthermore,
$\Delta_k(T_k)$ is the temperature-dependent superconducting energy gap and $\Delta_k(0)=\Delta_k=T_c 1.764 k_{\rm B}$, $T_c$ being the critical temperature of the superconductor, $f(\varepsilon,T_k)=\tanh(\varepsilon/2 k_BT_k)$,
$\Theta(x)$ is the Heaviside step function,
$k_B$ is the Boltzmann constant, $R_J$ is the junction normal-state resistance,
and $e$ is the electron charge.

On the other hand, for the phase-dependent component of the heat current one gets \cite{Maki,Guttman1,Guttman2,Sauls1,Sauls2} 
\begin{eqnarray}
\dot{Q}_{int}(T_{\rm hot},T_{\rm cold})=\frac{1}{e^2 R_J}\int^{\infty}_{0}d\varepsilon \varepsilon \mathcal{M}_{1}(\varepsilon,T_{\rm hot})\mathcal{M}_{2}(\varepsilon,T_{\rm cold})\nonumber \\ \times [f(\varepsilon,T_{\rm cold})-f(\varepsilon,T_{\rm hot})],
\label{qint}
\end{eqnarray}
where $\mathcal{M}_{k}(\varepsilon,T_k)=\frac{\Delta_{k}(T_k)}{\sqrt{\varepsilon^2-\Delta_{k}(T_k)^2}}   \Theta[\varepsilon^2-\Delta_{k}(T_k)^2]$ is the Cooper Pair BCS density of states in S$_{_k}$ at temperature $T_k$ \cite{Barone}. This term originates from those energy-carrying processes involving Cooper pair tunneling and recombination or destruction of Cooper pairs.
Since the phase difference between the annihilated and created pair is relevant in such a process this gives rise to the cos$\varphi$ contribution to the transferred heat.

%Note that both  $\dot{Q}_{qp}$ and $\dot{Q}_{int}$ vanish for $T_{\rm hot}= T_{\rm cold}$, while $\dot{Q}_{int}$ also vanishes when at least one of the superconductors is in the normal state, i.e., $\Delta_k(T_k)=0$.

Along this manuscript, we will neglect any contribution to thermal transport arising from lattice phonons. Moreover, we will assume the latter to be very well thermalized with the substrate phonons that reside at the bath temperature ($T_{\rm bath}$). This is usually the case in the kind of systems of our interest as the Kapitza resistance between a metallic thin film ($\sim 20-30$ nm) and the substrate is typically negligibly small at low temperatures \cite{Wellstood}.  It will be also convenient to provide with the expression for the energy exchanged between electrons and phonons in the superconductor \cite{Timofeev2}, 
\begin{eqnarray}
\dot{Q}_{\rm e-ph,S}(T_{\rm hot},T_{\rm bath})=-\frac{\Sigma_{S} \mathcal{V}_{S}}{96\zeta (5)k_B^5}   \int^\infty_{-\infty}dE E  \int^{\infty}_{-\infty}d\varepsilon \varepsilon^2 \textrm{sign}(\varepsilon) M_{E,E+\varepsilon} \nonumber\\ \times   \bigg[ \coth\Big(\frac{\varepsilon}{2k_B T_{\rm bath}}\Big)  \big(f(E,T_{\rm hot})-f(E+\varepsilon,T_{\rm hot})\big) \nonumber\\  -f(E,T_{\rm hot})f(E+\varepsilon,T_{\rm hot})+1  \bigg],
\label{eph}
\end{eqnarray}
where $M_{E,E'}(T_{\rm hot})=\mathcal{N}_1(E,T_{\rm hot})\mathcal{N}_1(E',T_{\rm hot})[1-\Delta_1^2(T_{\rm hot})/EE']$, $\Sigma_{S}$ is the electron-phonon coupling constant, and $\mathcal{V}_{S}$ is the volume of the superconducting electrode.

 The behavior of $\dot{Q}_{\rm tot}$ can be experimentally revealed by probing the electronic temperature of one of the superconducting electrodes or, alternatively, a normal metal electrode tunnel-connected to the junction. The latter option is more convenient from an experimental point of view as electron thermometry is typically simpler to be performed on normal metals \cite{RMP} and will be therefore considered here.

%%%%%%%%%%%%%%%%%%%%%%%%%%%%%%%%%%%%%%%%%%%%%%%%%%%%%%%%%%%%%%%%%%%%%%%%%%%%%%%%%%%%%%%%%%%%%%%%%%%%%%%%%%%%%%%%%%%%%%%%%%%%%%%%%%%%%%%%%%%%%%%%%%%%%%%%%%%%%%%%%%%%%%%%%%%%%%%%%%%%%%%%%%%%%%%%%%%%%%%%%%%%%%%%%%%%%%%%%%%%%%%%%%%%%%%%%%%%%%%%%%%%%%%%%%%%%%%%%%%%%%%%%%%%%%%%%%%%%%%%%%%%%%%%%%%%%%%%%%%%%%%%%%%%%%%%%%%%%%%%%%%%%%%%%%%%%%%%%%%%%%%%%%%%%%%%%%%%%%%%%%%%%%%%%%%%%%%%%%%%%%%%%%%%%%%%%%%%%%%%%%%%%%%%%%%%%%%%%%%%%%%%%%%%%%%%%%%%%%%%%%%%%%%%%%%%%%%%%%%%%%%%%%%%%%%%%%%%%%%%%%%%%%%%%%%%%%%%%%%%%%%%%%%%%%%%%%%%%%%%%%%%%%%%%%%%%%%%%%%%%%%%%%%%%%%%%%%%%%%%%%%%%%%%%%%%%%%%%%%%%%%%%%%%%%%%%%%%%%%%%%%%%%%%%%%%%%%%%%%%%%%%%%%%%%%%%%%%%%%%%%%%%%%%%%%%%%%%%%%%%%%%%%%%%%%%%%%%%%%%%%%%%%%%%%%%%%%%%%%%%%%%%%%%%%%%%%%%%%%%%%%%%%%%%%%%%%%%%%%%%%%%%%%%%%%%%%%%%%%%%%%%%%%%%%%%%%%%%%%%%%%%%%%%%%%%%%%%%%%%%%%%%%%%%%%%%%%%%%%%%%%%%%%%%%%%%%%%%%%%%%%%%%%%%%%%%%%%%%%%%%%%%%%%%%%%%%%%%%%%%%%%%%%%%%%%%%%%%%%%%%%%%%%%%%%%%%%%%%%%%%%%%%%%%%%%%%%%%%%%%%%%%%%%%%%%%%%%%%%%%%%%%%%%%%%%%%%%%%%%%%%%%%%%%%%%%%%%%%%%%%%%%%%%%%%%%%%%%%%%%%%%%%%%%%%%%%%%%%%%%%%%%%%%%%%%%%%%%%%%%%%%%%%%%%%%%%%%%%%%%%%%%%%%%%%%%%%%%%%%%%%%%%%%%%%%%%%%%%%%%%%%%%%%%%%%%%%%%%%%%%%%%%%%%%%%%%%%%%%%%%%%%%%%%%%%%%%%%%%%%%%%%%%%%%%%%%%%%%%%%%%%%%%%%%%%%%%%%%%%%%%%%%%%%%%%%%%%%%%%%%%%%%%%%%%%%%%%%%%%%%%%%%%%%%%%%%%%%%%%%%%%%%%%%%%%%%%%%%%%%%%%%%%%%%%%%%%%%%%%%%%%%%%%%%%%%%%%%%%%%%%%%%%%%%%%%%%%%%%%%%%%%%%%%%%%%%%%%%%%%%%%%%

\section{The Josephson heat modulator}
\label{sec1}

%In this section we recall the observation of heat interference based on the realization a of a phase-controlled superconducting heat-flux quantum modulator. 
We start by  theoretically investigating heat exchange between two normal metal electrodes kept at different temperatures and tunnel-coupled to each other through a thermal `modulator' \cite{GiazottoAPL12} in the form of a DC-SQUID. Heat transport in a similar system is subsequently studied experimentally and found to be phase dependent, in agreement with the original prediction \cite{Maki,Giazottoarxiv}.

\subsection{Theoretical considerations and layout design}
\label{subsec11}

\begin{figure}[t]
\centering
\includegraphics[width=0.9\textwidth]{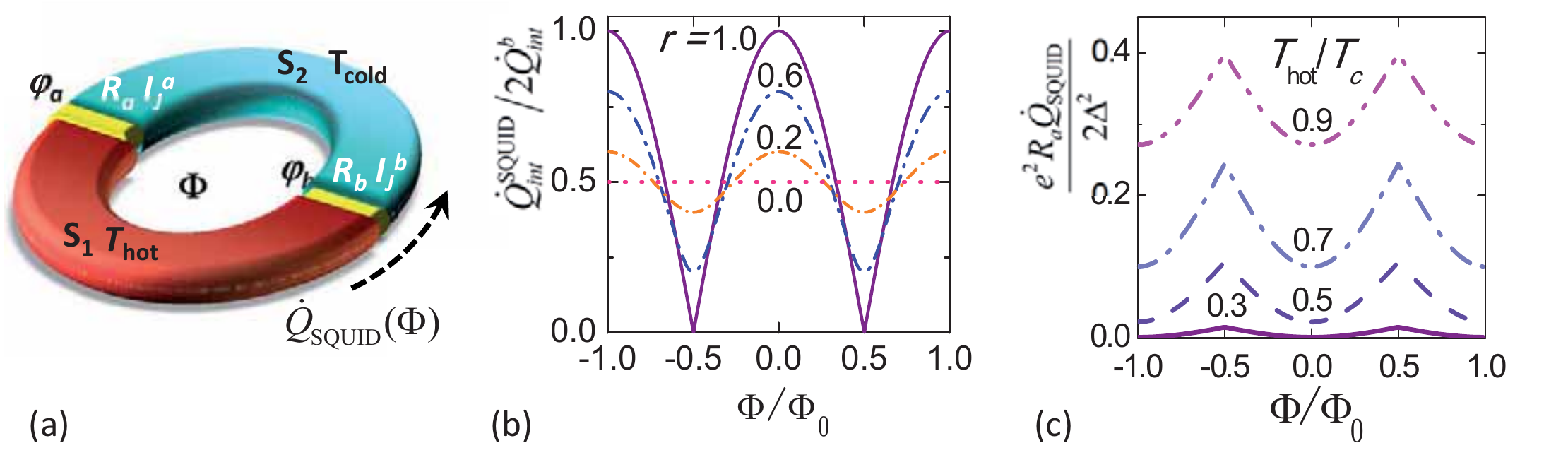}
\caption{(Color online) (a) Scheme of the proposed device.
Two superconductors S$_1$ and S$_2$ kept at temperature $T_{\rm hot}$ and $T_{\rm cold}$, respectively, are tunnel coupled so to implement a thermally-biased DC SQUID. %$\varphi_{a(b)}$  is the quantum phase difference over JJ$_{a(b)}$ having normal-state resistance $R_{a(b)}$, $\Phi$ is the applied magnetic flux threading the SQUID loop, and  $\dot{Q}_{\rm SQUID}(\Phi)$ is the total heat current flowing through the structure.  
(b) Interference heat current $\dot{Q}_{int}^{\rm SQUID}$ vs $\Phi$ calculated for a few values of $r=R_{b}/R_{a}$.
(c) Total heat current $\dot{Q}_{\rm SQUID}$ vs $\Phi$ calculated for a few values of $T_{\rm hot}$ at $T_{\rm cold}=0.1T_c$ assuming $r=1$. %(d) Average total heat current over one flux quantum $\left\langle \dot{Q}_{\rm SQUID}\right\rangle$ vs $T_{\rm hot}$ calculated for some values of $T_{\rm cold}$. The inset shows the total heat current modulation amplitude $\delta \dot{Q}_{\rm SQUID}$ vs $T_{\rm hot}$ calculated for the same $T_{\rm cold}$ values.
}
\label{Fig11} 
\end{figure}

The structure we envision  is sketched in Fig. \ref{Fig11}(a). It consists of a  DC SQUID composed of two superconductors S$_1$ and S$_2$ in thermal equilibrium kept at different temperatures $T_{\rm hot}$ and $T_{\rm cold}$, respectively. 
$R_{a(b)}$ and  $\varphi_{a(b)}$ denote the normal-state resistance and phase drop of JJ$_{a(b)}$, respectively.
In this specific case the total heat current given in Eq. (\ref{heatcurrent}) flowing from S$_1$ to S$_2$ becomes 
$\dot{Q}_{\rm SQUID}=\dot{Q}_{qp}^{\rm SQUID}(T_{\rm hot},T_{\rm cold})-\dot{Q}_{int}^{\rm SQUID}(T_{\rm hot},T_{\rm cold},\varphi_a,\varphi_b)$,
%\begin{equation} 
%\dot{Q}_{\rm SQUID}=\dot{Q}_{qp}^{\rm SQUID}(T_{\rm hot},T_{\rm cold})-\dot{Q}_{int}^{\rm SQUID}(T_{\rm hot},T_{\rm cold},\varphi_a,\varphi_b), 
%\label{qdot}
%\end{equation}
where   $\dot{Q}_{qp}^{\rm SQUID}=\dot{Q}_{qp}^a+\dot{Q}_{qp}^b$,
and 
$\dot{Q}_{int}^{\rm SQUID}=\dot{Q}_{int}^a \cos \varphi_a + \dot{Q}_{int}^b\cos\varphi_b$.
For definiteness, we assume that $T_{\rm hot}\geq T_{\rm cold}$ so that the SQUID is only biased with a temperature drop across the junctions, but the voltage across them vanishes.

%____________________________________________________________________________________________________________________________________________________________
%____________________________________________________________________________________________________________________________________________________________
%____________________________________________________________________________________________________________________________________________________________
%____________________________________________________________________________________________________________________________________________________________

By neglecting the geometric inductance of the ring it follows that $\varphi_a+\varphi_b+2\pi\Phi/\Phi_0=2n\pi$ where $\Phi$ is the applied magnetic flux through the loop, $n$ is an integer and $\Phi_0=2.067\times10^{-15}$ Wb is the flux quantum. 
For a given $\Phi$, the phases $\varphi_a$ and $\varphi_b$
are determined by the equation $i_{\rm J}^a \sin\varphi_a=i_{\rm J}^b \sin\varphi_b$, which describes conservation of the supercurrent circulating along the loop, where
$i_{\rm J}^{a(b)}\propto R_{a(b)}^{-1}$ is the Ambegaokar-Baratoff critical current \cite{Tinkham}  of junction JJ$_{a(b)}$. By defining $r=i_{\rm J}^a/i_{\rm J}^b=\dot{Q}_{int}^a/\dot{Q}_{int}^b=\dot{Q}_{qp}^a/\dot{Q}_{qp}^b=R_b/R_a$ (with $0\leq r\leq 1$) as the degree of asymmetry of the SQUID junctions one gets \cite{bo}
$\cos\varphi_a=r+\cos(2\pi \Phi/\Phi_0)/\sqrt{1+r^2+2r\cos(2\pi \Phi/\Phi_0)}$ and $\cos\varphi_b=1+r\cos(2\pi \Phi/\Phi_0)/\sqrt{1+r^2+2r\cos(2\pi \Phi/\Phi_0)}$.
Using the latter equations $\dot{Q}_{int}^{\rm SQUID}$ can be rewritten as $\dot{Q}_{int}^{\rm SQUID}=\dot{Q}_{int}^b\sqrt{1+r^2+2r\cos\left(2\pi\Phi/\Phi_0\right)}$,
which is analogous to the expression for the total Josephson \emph{electric} critical current in a DC SQUID with generic junctions asymmetry \cite{Clarke}. 
In particular, for  a symmetric SQUID ($r=1$) we get
$\dot{Q}_{int}^{\rm SQUID}=2\dot{Q}_{int}^b(T_{\rm hot},T_{\rm cold})\left|\cos (\pi\Phi/\Phi_0\right)|.$
%\begin{equation}
%\dot{Q}_{int}^{\rm SQUID}=2\dot{Q}_{int}^b(T_{\rm hot},T_{\rm cold})\left|\cos\left(\frac{\pi\Phi}{\Phi_0}\right)\right|.
%\label{symm}
%\end{equation}

We remark that, in this specific case, the phase-dependent component of the heat current always flows  in the direction opposite to the quasiparticle heat flow, i.e., from the cold to the hot electrode% [notice the minus sign in front of the equation for $\dot{Q}_{int}^{\rm SQUID}$] 
\cite{simmonds2012}.

%---------------------------------------------------------------------------------------------------------------

Figure \ref{Fig11} (b) shows $\dot{Q}_{int}^{\rm SQUID}$ vs $\Phi$ calculated for a few values of $r$ at generic temperatures $T_{\rm hot}$ and $T_{\rm cold}$ such that $T_{\rm cold}<T_{\rm hot}<T_c$. 
As it can be seen, $\dot{Q}_{int}^{\rm SQUID}$ is a periodic function of $\Phi$  maximized at integer values of $\Phi_0$. %, and is modulated between the maximum given by $\dot{Q}_{int}^b(1+r)$ and the minimum given by $\dot{Q}_{int}^b(1-r)$. For $r=1$, $\dot{Q}_{int}^{\rm SQUID}(\Phi)$ is thus modulated between $2\dot{Q}_{int}^b$ and 0 [see Eq. (\ref{symm})]. 
By increasing $r$ leads to a suppression of the modulation amplitude combined with a reduction of the average value of the heat current. Eventually, the modulation amplitude is totally suppressed for $r=0$, as only one junction is driving heat flow through the SQUID. 
Therefore high junctions symmetry is desired to maximize heat current modulation in the device. For that reason, in the following we will restrict our calculations to the case $r=1$. 
Figure \ref{Fig11}(c) shows the total heat current $\dot{Q}_{\rm SQUID}$ vs $\Phi$  at $T_{\rm cold}=0.1T_c$ for a few values of $T_{\rm hot}$. 
$\dot{Q}_{\rm SQUID}$ is $\Phi_0$-periodic and minimized for integer values of $\Phi_0$. %At the lowest $T_{\rm hot}$ the total heat current is small, as $\dot{Q}_{qp}^{\rm SQUID}$ and $\dot{Q}_{int}^{\rm SQUID}$ are comparable in magnitude. By increasing $T_{\rm hot}$ leads to the enhancement of both the average heat current and modulation amplitude which originate from larger temperature drop across the junctions. Further enhancement of $T_{\rm hot}$ leads to a suppression of the heat current modulation amplitude which disappears at $T_{\rm hot}=T_c$ when S$_1$ is driven into the normal state.

\subsection{Experimental realization}
\label{subsec12}

\begin{figure}[t]
\centering
\includegraphics[width=.75\textwidth]{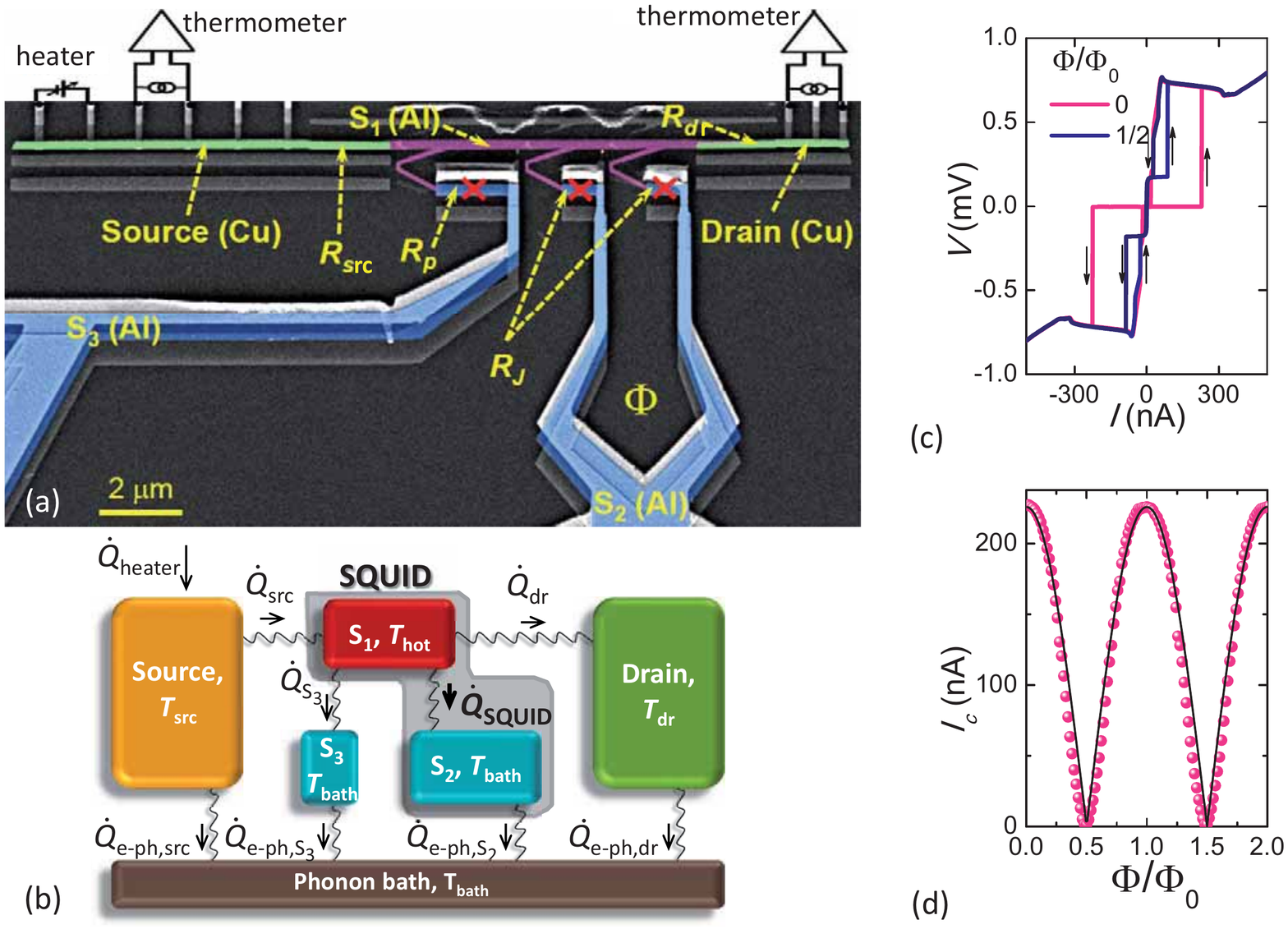}
\caption{(Color online) (a)  Scanning electron micrograph of the heat interferometer. %Cu source and drain electrodes share the contact through AlO$_x$ tunnel barriers with an Al island (S$_1$)  defining one branch of a DC-SQUID. The other branch of the SQUID (S$_2$) extends into a large volume lead to insure proper thermalization of its quasiparticles at the bath temperature $T_{\rm bath}$. 
SQUID junctions are marked by crosses. 
%S$_1$ is also in contact with an Al tunnel probe (S$_3$) enabling independent characterization of the SQUID.NIS junctions made of Al  are used as heaters and thermometers. See text for details. 
(b) Idealized thermal diagram accounting for our setup. %S$_1$ exchanges energy at power $\dot{Q}_{\rm src}$ and $\dot{Q}_{\rm dr}$ due to quasiparticle heat conduction with source and drain, respectively, at power $\dot{Q}_{\rm SQUID}$ with S$_2$ and $\dot{Q}_{S_3}$ with S$_3$. S$_2$ and S$_3$ are assumed to reside at $T_{\rm bath}$. Drain electrons exchange energy at power $\dot{Q}_{\rm dr}$ with S$_1$, and at power $\dot{Q}_{\rm e-ph,dr}$ with lattice phonons thermalized at bath temperature.
 Arrows indicate the direction of heat flows for $T_{\rm bath}<T_{\rm dr}<T_{\rm hot}<T_{\rm src}$.
(c) SQUID voltage ($V$) versus current ($I$) characteristics at two representative values of the applied flux and 
%$\Phi_0$ corresponds to an applied field $B=\Phi_0/A\approx 1$ Oe, where $A\approx 19.6\,\mu$m$^2$ is the ring area. 
(d) $\Phi$-dependent experimental pattern of the SQUID critical current $I_c$ along with the theoretical curve (solid line) %for a DC-SQUID assuming $\sim 0.3\%$ asymmetry between the critical currents of the two junctions.
 taken at 240 mK.
} \label{Fig13}
\end{figure}

The experimental implementation of our heat interferometer is shown in Fig. \ref{Fig13}(a). The structure consists of a source and drain copper (Cu) electrodes tunnel-coupled to a superconducting aluminum (Al) island  defining one half of a DC-SQUID (S$_1$).  The other half of the SQUID (S$_2$) extends into a large volume lead to insure proper thermalization of its quasiparticles at the bath temperature $T_{\rm bath}$.
S$_1$ is also contacted by an extra Al probe (S$_3$) via a tunnel junction, enabling independent characterization of the SQUID.
Both source and drain are tunnel-coupled to a few external Al probes so to realize normal metal-insulator-superconductor (NIS) junctions, which allow Joule heating and thermometry \cite{RMP}\footnote{ Source, drain and S$_3$ junctions normal-state resistances are $R_{\rm src}\simeq 1.5$ k$\Omega$, $R_{\rm dr}\simeq 1$ k$\Omega$ and $R_{\rm p}\sim 0.55$ k$\Omega$ , respectively, whereas the resistance of each SQUID junction is $R_{\rm J}\simeq 1.3$ k$\Omega$. The ring area is $\sim 19.6\,\mu$m$^2$. Finally, NIS thermometers exhibit $\sim 25$ k$\Omega$ normal-state resistance each.}.

%The structure  has been fabricated with e-beam lithography and three-angle shadow-mask evaporation of metals onto an oxidized Si wafer through a suspended resist mask. In the e-beam evaporator, the chip is initially tilted at an angle of $28^{\circ}$, and 20 nm of Al are deposited to realize S$_2$ and S$_3$. The sample is then exposed to 380 mTorr  of O$_2$ for 4.5 minutes to form the SQUID tunnel barriers after which it is tilted to $0^{\circ}$ for the deposition of 25 nm of Al forming S$_1$ as well as heaters and thermometers probes. The chip is subsequently exposed to 800 mTorr of O$_2$ for 4.5 minutes to form heaters, thermometers, source and drain tunnel junctions. Finally, 30 nm of Cu are deposited at $42^{\circ}$ to realize source and drain. 
%____________________________________________________________________________________________________________________________________________________________
%____________________________________________________________________________________________________________________________________________________________
%____________________________________________________________________________________________________________________________________________________________
%____________________________________________________________________________________________________________________________________________________________

Below the critical temperature of Al ($\simeq 1.4$ K) Josephson coupling allows dissipationless charge transport through the SQUID. The SQUID voltage-current characteristics at $240$ mK for two representative magnetic-flux values are shown in Fig. \ref{Fig13}(c). 
In particular, a well-defined Josephson current with maximum amplitude of $\simeq226$ nA is observed at $\Phi=0$.
The magnetic-flux pattern of the SQUID critical current $I_c$ along with the theoretical prediction \cite{Clarke} is displayed in Fig. \ref{Fig13}(d), and shows a nearly-complete supercurrent modulation, which confirms the good symmetry of the SQUID.

Thermal transport and, therefore, heat interference in the structure, arises from heating electrons in the source above lattice temperature ($T_{\rm bath}$) so to elevate the quasiparticles temperature in S$_1$ ($T_{\rm hot}$).  As the second half of the SQUID (S$_2$) is well thermalized at $T_{\rm bath}$ a temperature gradient arises across the SQUID. $\dot{Q}_{\rm SQUID}$ will thus manifest itself leading to a $\Phi_0$-periodic modulation of drain electron temperature ($T_{\rm dr})$.

%____________________________________________________________________________________________________________________________________________________________
%____________________________________________________________________________________________________________________________________________________________
\begin{figure}[t]
\centering
\includegraphics[width=0.8\textwidth]{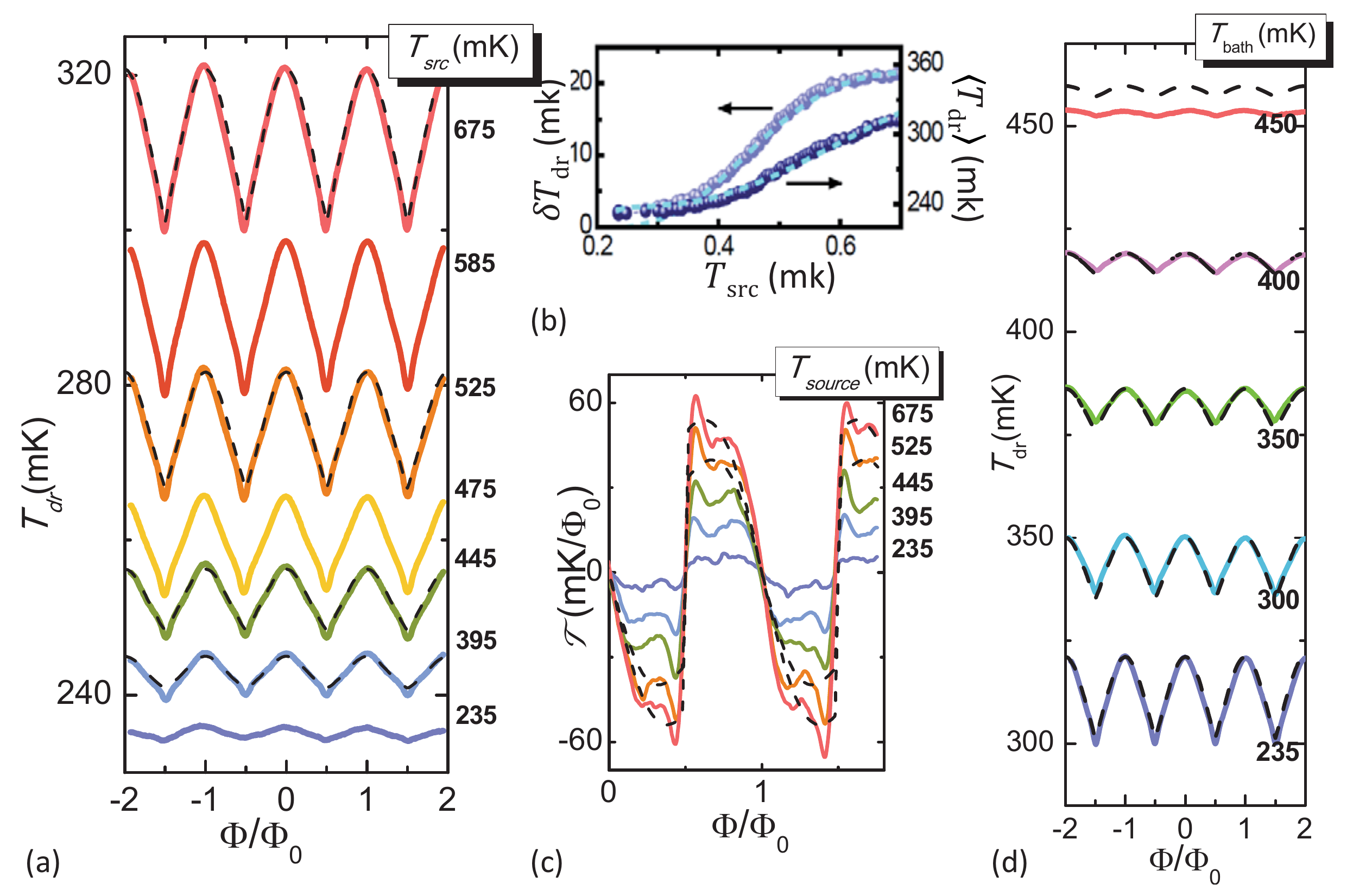}
\caption{(Color online) (a) Flux modulation of $T_{\rm dr}$ measured for several $T_{\rm src}$ values.
(b) Modulation amplitude $\delta T_{\rm dr}$  and average temperature $\left\langle T_{\rm dr}\right\rangle$  vs. $T_{\rm src}$.
(c) Flux-to-temperature transfer function $\mathcal{T}\equiv \partial T_{\rm dr}/\partial \Phi$ versus $\Phi$ measured at a few selected $T_{\rm src}$.  (d)
Flux modulation of $T_{\rm dr}$ recorded at different $T_{\rm bath}$ and $T_{\rm src}\approx 700$ mK. 
%From bottom to top, data correspond to $T_{\rm src}=675$ mK, $700$ mK, $690$ mK, $700$ mK, $700$mK.
Dashed lines in all panels are the theoretical results obtained from the thermal model (see text).   
}\label{Fig14} 
\end{figure}
%____________________________________________________________________________________________________________________________________________________________
%____________________________________________________________________________________________________________________________________________________________

Investigation of heat transport in our system is performed by stabilizing first the cryostat temperature at a desired $T_{\rm bath}$ and heating the source up to a given temperature $T_{\rm src}$. For this purpose, one pair of NIS junctions in the source is operated as a heater whereas a second pair is used to measure $T_{\rm src}$ by applying a small DC bias current and recording the  corresponding temperature-dependent voltage drop \cite{RMP,Nahum}. 
Analogously, another pair of NIS junctions is used to probe the electronic drain temperature ($T_{\rm dr}$) against a slowly-sweeping external magnetic flux.

Figure \ref{Fig14}(a) shows $T_{\rm dr}$ against $\Phi$ measured at 235 mK for increasing values of $T_{\rm src}$. 
Notably, $T_{\rm dr}$ is $\Phi_0$-periodic in $\Phi$, as the Josephson critical current [see Fig. \ref{Fig13}(d)], and  maximized at integer values of $\Phi_0$. 
As we shall argue, such a temperature modulation is of coherent nature, and stems from magnetic flux-control of $\dot{Q}_{\rm SQUID}$, which is a hallmark of the Josephson effect. 
A monotonic enhancement of the average drain temperature over one flux quantum, $\left\langle T_{\rm dr}\right\rangle$, is observed when raising $T_{\rm src}$, which follows from increased heat flow across the structure. 
On the other hand, the modulation amplitude  $\delta T_{\rm dr}$, defined as the difference between the maximum and minimum values of $T_{\rm dr}$, turns out to initially increase and then tends to saturate at larger $T_{\rm src}$.
In particular, $\delta T_{\rm dr}$ up to $\sim 21$ mK is observed corresponding to $\sim 9\%$ of relative modulation amplitude at 235 mK.
%$T_{\rm dr}$ modulation is observed also without intentional source heating, and might be related to a parasitic power ($\sim1-5$ fW) in the structure.    
The full $T_{\rm src}$-dependence of $\left\langle T_{\rm dr}\right\rangle$ and $\delta T_{\rm dr}$ are displayed in Fig. \ref{Fig14}(b)  and confirm the above described behavior. 
%We stress that heat interference manifests itself only owing to the existence of a finite temperature bias across the SQUID. Any voltage drop occurring at the JJs makes the phase-coherent component of  $\dot{Q}_{\rm SQUID}$ time-dependent thus  not contributing to time-averaged heat transport \cite{Maki,Guttman1,Guttman2}.

A relevant figure of merit of the heat interferometer is represented by the flux-to-temperature transfer coefficient, $\mathcal{T}\equiv \partial T_{\rm dr}/\partial \Phi$, shown in Fig. \ref{Fig14}(c)  versus $\Phi$ for a few selected $T_{\rm src}$. 
It turns out that $|\mathcal{T}|$ exceeding $60$ mK/$\Phi_0$ is obtained at $675$ mK.  %Larger values might be obtained by lowering $T_{\rm bath}$ and by further optimizing the structure design.

The dependence on bath temperature is shown in Fig. \ref{Fig14}(d), which displays $T_{\rm dr}(\Phi)$  at a few increasing $T_{\rm bath}$ for $T_{\rm src}$ set around 700 mK. Besides leading to a monotonic enhancement of $\left\langle T_{\rm dr} \right\rangle$, by increasing $T_{\rm bath}$ yields suppression of $\delta T_{\rm dr}$ and smearing of $T_{\rm dr}(\Phi)$, which can be mainly ascribed to the enhancement of electron-phonon coupling in the drain \cite{RMP} as well as to the influence of 
thermal broadening. 
$\delta T_{\rm dr}\sim 2.5$ mK is still observable at 450 mK, whereas the modulation disappears for $T_{\rm bath} \geq 500$ mK. 
We emphasize that the latter is substantially smaller than the temperature setting the disappearance of the Josephson effect in the SQUID ($\simeq 1.4$ K). 
%Figures \ref{Fig16}(c)-(e)  show  $\left\langle T_{\rm dr} \right\rangle$, $\delta T_{\rm dr}$, and the maximum of $|\mathcal{T}|$ versus $T_{\rm src}$, respectively, recorded at the same $T_{\rm bath}$ as in panel (b), along with the lines obtained from thermal model. Each of these quantities displays similar qualitative behavior at different bath temperatures, and a smoothing of their characteristics is observed by increasing $T_{\rm bath}$.  This is consistent with the picture provided by the model which captures the main features of the experimental data. These results confirm what originally predicted almost fifty years ago \cite{Maki}, i.e., the existence of a phase-dependent component in the heat current flowing  through a temperature-biased Josephson tunnel junction. 

To account for our observations we have elaborated a thermal model sketched
in Fig. \ref{Fig13}(b). We assume S$_1$ to exchange heat at power $\dot{Q}_{\rm src}$ and $\dot{Q}_{\rm dr}$ due to quasiparticle heat conduction with source and drain, respectively, at power $\dot{Q}_{\rm SQUID}$ with S$_2$ and $\dot{Q}_{\rm S_3}$ with S$_3$.
Both S$_2$  and S$_3$ are assumed to be thermalized at $T_{\rm bath}$. 
Furthermore, drain electrons exchange energy at power $\dot{Q}_{\rm dr}$ with S$_1$, and at power  $\dot{Q}_{\rm e-ph,dr}$ with lattice phonons residing at $T_{\rm bath}$ \cite{RMP,Roukes}. 
The thermal steady-state of the system may be described by the energy-balance equations
 $-\dot{Q}_{\rm src}+\dot{Q}_{\rm S_3}+\dot{Q}_{\rm SQUID}(\Phi)+\dot{Q}_{\rm dr}=0$ and $-\dot{Q}_{\rm dr}+\dot{Q}_{\rm e-ph,dr}=0$, where first equation accounts for thermal budget in S$_1$, while the second one describes heat exchange in the drain. 
$T_{\rm hot}$ and $T_{\rm dr}$ can be determined under given conditions by numerically solving the aforementioned equations\footnote{$\dot{Q}_{S_3}=\dot{Q}_{tot}(T_{\rm hot},T_{\rm cold},0)$ [see Eq. (\ref{heatcurrent})] substituting $R_{\rm J}$ for $R_{\rm p}$, $\mathcal{N}_2 (\varepsilon,T_{\rm cold})$ for $\mathcal{N}_3 (\varepsilon,T_{\rm cold})=\mathcal{N}_2 (\varepsilon,T_{\rm cold})$ and $\mathcal{M}_2 (\varepsilon,T_{\rm cold})$ for $\mathcal{M}_3 (\varepsilon,T_{\rm cold})=\mathcal{M}_2 (\varepsilon,T_{\rm cold})$.
Furthermore, $\dot{Q}_{\rm src}$ can be obtained from Eq. (\ref{qqp}) by doing $\dot{Q}_{\rm src}=\dot{Q}_{qp}(T_{\rm hot}, T_{\rm src})$ with $\mathcal{N}_2=1$ and substituting $R_{\rm J}$ for $R_{\rm src}$,  $\dot{Q}_{\rm dr}$ can be also obtained from Eq. (\ref{qqp}) by setting $\dot{Q}_{\rm dr}=-\dot{Q}_{qp}(T_{\rm hot}, T_{\rm dr})$ with $\mathcal{N}_2=1$ and substituting $R_{\rm J}$ for $R_{\rm dr}$. For the normal metal $\dot{Q}_{\rm e-ph,dr}=\Sigma_{\rm dr} \mathcal{V}_{\rm dr}(T_{\rm dr}^5-T_{\rm bath}^5)$, where $\Sigma_{\rm dr} \simeq 3\times 10^9$ WK$^{-5}$m$^{-3}$ is the electron-phonon coupling constant for Cu \cite{RMP}, and $\mathcal{V}_{\rm dr}\simeq 2\times 10^{-20}$ m$^{3}$ is drain volume.}.
The model neglects heat exchange with photons due to mismatched impedance \cite{Cleland,meschkenature,Timofeev}, electron-phonon coupling in S$_1$ owing to its reduced volume and low experimental $T_{\rm bath}$ \cite{Timofeev2}, as well as phonon heat current \cite{Maki}. Theoretical curves have been obtained by setting the structure parameters as extracted from the experiment, and varying $R_{\rm p}$ between  $\sim 100\%$ and $\sim 125\%$ to match measured data. Results from the thermal model are shown in Fig. \ref{Fig14} (dashed lines).  Although idealized, this model provides reasonable agreement with our observations, and grasps the relevant physical picture at the origin of heat interference in our system.

%In the energy-balance equations, $\dot{Q}_{S_3}=\dot{Q}_{qp}^{S_3}-\dot{Q}_{int}^{S_3}$ \cite{Maki,Guttman2,GiazottoAPL12,Guttman1, Sauls1,Sauls2},
%$\dot{Q}_{qp}^{S_3}=\frac{2}{e^2 R_{\rm p}}\int^{\infty}_{0} d\varepsilon \varepsilon \mathcal{N}_1 (\varepsilon)\mathcal{N}_3 (\varepsilon)[f_1(\varepsilon)-f_3(\varepsilon)]$,
%$\dot{Q}_{int}^{S_3}=\frac{2}{e^2 R_{\rm p}}\int^{\infty}_{0} d\varepsilon \varepsilon \mathcal{M}_1 (\varepsilon)\mathcal{M}_3 (\varepsilon)[f_1(\varepsilon)-f_3(\varepsilon)]$, $\mathcal{N}_3 (\varepsilon)=\mathcal{N}_2 (\varepsilon)$, $\mathcal{M}_3 (\varepsilon)=\mathcal{M}_2 (\varepsilon)$, and $f_3(\varepsilon)=f_2(\varepsilon)$.  
%Furthermore, $\dot{Q}_{\rm src}=\frac{2}{e^2 R_{\rm src}}\int^{\infty}_{0} d\varepsilon \varepsilon \mathcal{N}_1 (\varepsilon)[f(\varepsilon,T_{\rm src})-f(\varepsilon,T_{\rm hot})]$, $\dot{Q}_{\rm dr}=\frac{2}{e^2 R_{\rm dr}}\int^{\infty}_{0} d\varepsilon \varepsilon \mathcal{N}_1 (\varepsilon)[f(\varepsilon,T_{\rm hot})-f(\varepsilon,T_{\rm dr})]$ \cite{RMP}, $\dot{Q}_{\rm e-ph,dr}=\Sigma_{\rm dr} \mathcal{V}_{\rm dr}(T_{\rm dr}^5-T_{\rm bath}^5)$ \cite{RMP,Roukes}, $\Sigma_{\rm dr} \simeq 3\times 10^9$ WK$^{-5}$m$^{-3}$ is the electron-phonon coupling constant for Cu \cite{RMP}, and $\mathcal{V}_{\rm dr}\simeq 2\times 10^{-20}$ m$^{3}$ is drain volume.

\section{Double-ring heat modulator}
\label{sec2}

In this section we envision and theoretically analyze a double-loop heat interferometer \cite{martinez2012}. The latter allows to finely tune and fully balance the heat flux flowing through the device making it an improved \emph{heat transistor}. Additionally, it is much more robust against fabrication deficiencies such as differences between the normal-state resistances of the JJs. Thanks to the application of two independent magnetic fluxes, even a quite asymmetric device is able of providing a much more robust temperature modulation compared to the simple single-loop heat interferometer or its complete annihilation.

\subsection{Theoretical considerations and layout design}
\label{subsec2}

\begin{figure}[t]
\centering
\includegraphics[width=0.9\columnwidth]{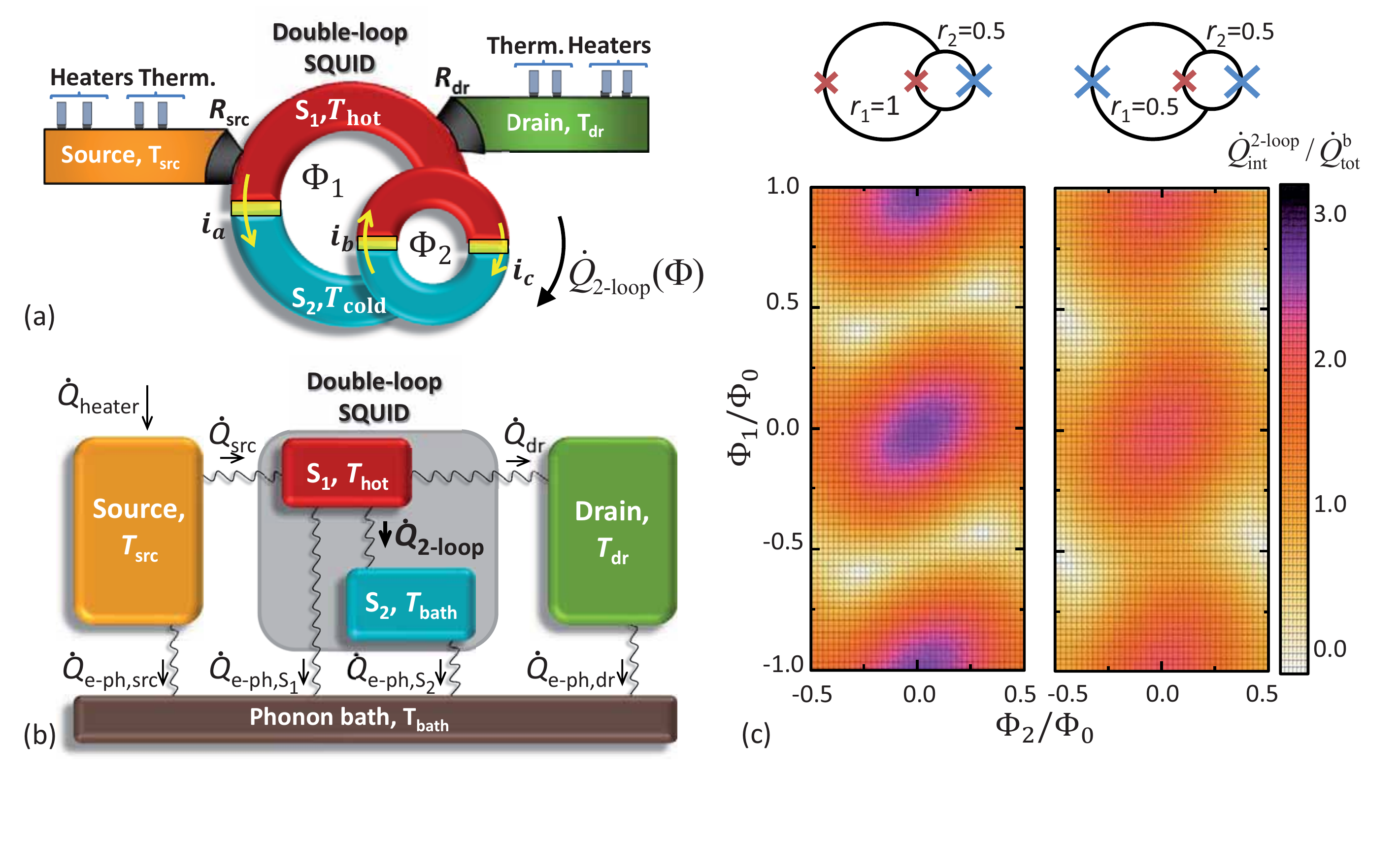}
\caption{(Color online) (a) Three parallel JJs define a double-loop heat interferometer. The temperature in S$_1$ is risen up to $ T_{\rm hot}\geq T_{\rm cold}$, which yields a steady-state heat current $\dot{Q}_{\rm 2-loop}$. Source and drain normal-metal electrodes are tunnel-coupled to S$_1$. Superconducting tunnel junctions operated as heaters and thermometers are connected to source and drain.  (b)  Idealized thermal diagram accounting for our setup. %S$_1$ exchanges energy at power $\dot{Q}_{\rm src}$ and $\dot{Q}_{\rm dr}$ due to quasiparticle heat conduction with source and drain, respectively, and at power $\dot{Q}_{\rm 2-loop}$ with S$_2$. S$_2$  resides at $T_{\rm bath}$. Drain electrons exchange energy at power $\dot{Q}_{\rm dr}$ with S$_1$, and at power $\dot{Q}_{\rm e-ph,dr}$ with lattice phonons thermalized at bath temperature. 
Arrows indicate the direction of heat flows for $T_{\rm bath}<T_{\rm dr}<T_{\rm hot}<T_{\rm src}$. (c) Density plots showing $\dot{Q}_{int}^{\rm 2-loop}$ as a function of the magnetic flux on the main and the control loop, $\Phi_1$ and $\Phi_2$, respectively. Two representative cases have been considered and are  schematized on the top part of each plot. %The maximum and minimum values of $\dot{Q}_{int}^{\rm 2-loop}$ change perceptibly when reducing the symmetry of the double-loop. 
}
\label{Fig21}
\end{figure}

Our thermal circuit consists of two superconductors S$_1$ and S$_2$, weak linked forming a double-loop interrupted by three parallel JJs [see Fig. \ref{Fig21}(a)]. 
$R_k$ and  $\varphi_k$ denote the normal-state resistance and phase drop of junction $k$, with $k=a,b,c$.
This structure behaves as a conventional SQUID pierced by a magnetic flux $\Phi_1$ in which one of the junctions has been replaced by a DC SQUID. The characteristics of this second ``junction'' can be tuned thanks to the application of a control magnetic flux $\Phi_2$. The system is temperature biased by setting the temperature in S$_1$ to be $T_{\rm hot}\geq T_{\rm cold}$, $T_{\rm cold}$ being the temperature in S$_2$. Furthermore, the voltage drop across the whole structure is set to zero. Under these circumstances, a thermal gradient arises across the junctions and a stationary heat current $\dot{Q}_{\rm 2-loop}$ will flow from S$_1$ to S$_2$, which are in steady-state thermal equilibrium \cite{Maki,Guttman1,Guttman2,Sauls1,Sauls2}. %As it was argued in Sec. \ref{sec02}, $ \dot{Q}_{\rm 2-loop}$ results from the sum of two terms, 
%\begin{equation}
%\dot{Q}_{\rm 2-loop} =  \dot{Q}_{qp}^{\rm 2-loop}(T_{\rm hot}, T_{\rm cold})-\dot{Q}_{int}^{\rm 2-loop}(T_{\rm hot}, T_{\rm cold},\varphi_a,\varphi_b,\varphi_c).
%\label{q2loop}
%\end{equation}

%____________________________________________________________________________________________________________________________________________________________
%____________________________________________________________________________________________________________________________________________________________
%____________________________________________________________________________________________________________________________________________________________
%____________________________________________________________________________________________________________________________________________________________

We shall concentrate on the phase-dependent heat current only given by $\dot{Q}_{int}^{\rm 2-loop} = \sum_{k=a,b,c} \dot{Q}_{int}^k(T_{\rm hot}, T_{\rm cold}) \cos \varphi_k$.
On the one hand, neglecting the self-induced flux in the loops,  the fluxoid quantization on both rings imposes $ \varphi_a + \varphi_b + 2 \pi \Phi_1/\Phi_0 = 2 n \pi$ and $\varphi_b + \varphi_c + 2 \pi \Phi_2/\Phi_0 = 2 m \pi$, where $n$ and $m$ are integers. %In these expressions we have neglected the geometric inductance of the SQUID which means neglecting the self-induced flux in the loops. %In a practical situation, these loops can be designed so to exhibit inductances $L_j$ of the order of a few tens of pH where $j = 1$, $2$ refers to the loop pierced by $\Phi_1$ and $\Phi_2$, respectively. Such geometry ensures a sufficiently good coupling between the SQUID loops and the additional control coils that couple $\Phi_1$ and $\Phi_2$ while providing reasonably low self-inductances.  Assuming  that each JJ $k$, with $k=a, b, c$, attains a maximal critical current $i^k_{\rm J}$ of the order of   $10^2$  nA one obtains a total screening parameter $\beta = 2 L_j i^k_{\rm J}/ \Phi_0 \sim 10^{-3} \ll 1$ allowing us to neglect the SQUID inductance. On the other hand, the consideration of finite $L_j$ makes the deduction of analytical expressions impossible and does not contribute to the understanding of the essential physics in our device. 
The conservation of the circulating supercurrent in both loops, on the other hand, imposes $ i^a_{\rm J} \sin \varphi_a =  i^b_{\rm J} \sin\varphi_b - i^c_{\rm J} \sin\varphi_c$, where $i^k_{\rm J} \propto R^{-1}_k$. In writing the previous equation we have established a given current sign convention [see yellow arrows in Fig. \ref{Fig21}(a)]. %We have done so in order to preserve consistency with the one-loop heat-flux modulator proposed in Ref. \citenum{GiazottoAPL12}.
Furthermore we define  $r_1 = i^a_{\rm J}/i^b_{\rm J} = \dot{Q}_{int}^a/\dot{Q}_{int}^b = R_b/R_a \geq0$, and $r_2 = i^c_{\rm J}/i^b_{\rm J} = \dot{Q}_{int}^c/\dot{Q}_{int}^b = R_b/R_c $. %$r_2$ can vary between $0 \leq r_2 \leq 1$ since setting $r_2 >1$ is equivalent to exchange the roles of $R_b$ and $R_c$. 
Combining the aforementioned conditions and using simple trigonometric relations one gets 
$  \cos \varphi_a = (r_1 + \alpha + r_2 \gamma)/ \delta$, $\cos \varphi_b =  (1 + r_1 \alpha + r_2 \beta)/ \delta$ and $\cos \varphi_c =  (r_2 + \beta + r_1 \gamma)/ \delta $, 
where $\delta=\sqrt{1+r^2_1+r^2_2+2r_1 \alpha+2r_2 \beta +2r_1 r_2 \gamma}$, $\alpha= \cos (2 \pi \Phi_1/\Phi_0)$, $\beta= \cos (2 \pi \Phi_2/\Phi_0)$ and $\gamma=\cos [ 2 \pi(\Phi_1 -  \Phi_2)/\Phi_0]$. %The choice of the signs of Eqs. (\ref{cosenos}) obeys the requirement of minimizing the free energy ($E_{\rm J}$) of the whole system. The latter can be written as $E_{\rm J} = 3E_{J,0}- E_{J,0}\sum _{k=a,b,c}i^k_{\rm J} \cos \varphi_k$, where $E_{J,0}=\Phi_0/ 2 \pi$ \cite{Tinkham}. By inserting Eqs. (\ref{cosenos}) into the previous expression it can be easily seen that the minimum of $E_{\rm J}$ is obtained for the solutions with positive signs. These, inserted into Eq. (\ref{eq1}), yield finally the following expression for $\dot{Q}_{int}^{\rm 2-loop}$,
Finally, one gets $  \dot{Q}_{int}^{\rm 2-loop}=\dot{Q}^{b}_{int}\delta$. 
%We note that, if we set $r_2=0$, i.e., $R_c\rightarrow\infty$, one recovers the expression of Eq. (\ref{totalQ}) corresponding to the single-loop heat interferometer. 
In Fig. \ref{Fig21}(c) we show two density plots of $\dot{Q}_{int}^{\rm 2-loop}$ vs. $\Phi_1$ and $\Phi_2$ for two representative cases. In general, the maximum of $\dot{Q}_{int}^{\rm 2-loop}$ is always reduced for the cases in which one resistance is different from the others. 

%Let us analyze in more detail what happens with $\delta \dot{Q}_{int}^{\rm 2-loop}$. If $r_1 \leq1$ , i.e., $R_a\leq R_b$, we find that the maximum amplitude of oscillation at fixed $\Phi_2$ is given by $\delta \dot{Q}_{int}^{\rm 2-loop}=  2\dot{Q}_{int}^b r_1 $ for whatever value of $r_2$, i.e., for whatever value of $R_c$. This is to say, $\delta \dot{Q}_{int}^{\rm 2-loop}$ is independent of the degree of asymmetry of the control loop and increases linearly by increasing the symmetry on the main loop. On the other hand, if  $r_1> 1$ the dependence of $\delta \dot{Q}_{int}^{\rm 2-loop}$ at fixed $\Phi_2$ is more complicated. For $r_1 - r_2 \leq 1$, $\delta \dot{Q}_{int}^{\rm 2-loop}$ is independent from the degree of symmetry in both loops reaching it maximum value, that, in this case, is given by $\delta \dot{Q}_{int}^{\rm 2-loop}=  2\dot{Q}_{int}^a$.  Finally, for $r_1 - r_2 > 1$, $\delta \dot{Q}_{int}^{\rm 2-loop}$ decreases with $r_2$. %These regimes are summarized in Fig. \ref{Fig4}(a).

The straightest choice is $r_1=r_2=1$. Although being the most simple configuration, it enables us to infer most of the characteristics of our thermal interferometer. When analyzing the behavior of $ \dot{Q}_{int}^{\rm 2-loop}$ vs. $\Phi_1$ for different $\Phi_2$ values one can distinguish between two regimes; In the first one, defined by $0 \leq \Phi_2 < \Phi_0 /3$ [Fig. \ref{Fig22}(a)], the mean value and the shape of the curves evolves whereas the amplitude of the modulation, denoted $\delta \dot{Q}_{int}^{\rm 2-loop}$, holds unchanged  as we increase the amplitude of the control flux $\Phi_2$. At $\Phi_2 = \Phi_0 / 3$  one gets $\dot{Q}_{int}^{\rm 2-loop}=2 J^{b}_{int} | \cos[ \pi (\Phi_1/\Phi_0-1/6)] |$. This is to say, apart from a small shift equal to $\Phi_0 /6 $, one recovers the same dependence on $\Phi_1$ obtained for the symmetric single-loop heat interferometer.  In the second regime, covered by $\Phi_0/3\leq\Phi_2\leq\Phi_0 /2$ [Fig. \ref{Fig22}(b)], $\delta \dot{Q}_{int}^{\rm 2-loop}$ decreases linearly with $\Phi_2$ whereas the mean value of $ \dot{Q}_{int}^{\rm 2-loop}$ remains constant. Furthermore, at $\Phi_2 = \Phi_0 /2 $, the modulation disappears completely and $\dot{Q}_{int}^{\rm 2-loop}$ becomes independent of $\Phi_1$.  The aforementioned characteristics are emphasized in Figs. \ref{Fig22}(c) and \ref{Fig22}(d) where we plot the transfer function, $\mathcal{T} _{\rm J} = \partial \dot{Q}_{int}^{\rm 2-loop} / \partial \Phi_1 $, for both regimes.

Similar curves are obtained in the more general case for which $r_1\neq r_2\neq 1$. Moreover, if the condition $i_a - i_c \leq i_b \leq i_a + i_c$ is satisfied,  it can be demonstrated that $ \dot{Q}_{int}^{\rm 2-loop}$ can be written as a function of $ | \cos[ \pi (\Phi_1/\Phi_0-\theta)] | $ where $\theta$ is a shift in $\Phi_1$. This is to say it is still possible to suppress completely $ \dot{Q}_{int}^{\rm 2-loop}$. Unlikely to a single-loop heat interferometer, even a quite asymmetric double-loop structure offers therefore the possibility of maximizing or suppressing completely $ \dot{Q}_{int}^{\rm 2-loop}$ through an appropriate choice of $\Phi_2$. %The minimum of $ \dot{Q}_{int}^{\rm 2-loop}$ can be easily determined from Eq. (\ref{blabli}). By requesting $ \partial \dot{Q}_{int}^{\rm 2-loop} / \partial \Phi_1 = 0$ and $ \partial \dot{Q}_{int}^{\rm 2-loop} / \partial \Phi_2 = 0$, and by summing the resulting equations we get that $r_1  \sin  (2 \pi \Phi_1/\Phi_0) + r_2 \sin  (2 \pi \Phi_2/\Phi_0) =0$. This equation gives us the family of values of $\Phi_1$ and $\Phi_2$ that maximize or minimize $  \dot{Q}_{int}^{\rm 2-loop}$ for any given $r_1$ and $r_2$. For instance, if $\Phi_1 = \Phi_0 /2$, there are only two solutions, $\Phi_2 =0$ and $\Phi_2 = \Phi_0 /2$, that satisfy this condition and, in addition, correspond to a minimum. Inserting these solutions into Eq. (\ref{blabli}) and imposing $\dot{Q}_{int}^{\rm 2-loop} = 0$ we obtain $r_1 - r_2 = 1$ or $r_1 + r_2 =1$, respectively. These equations define two straight lines in the $r_1 r_2$ plane. Within this area there exists at least one value of $\Phi_2$ that enables us to write $ \dot{Q}_{int}^{\rm 2-loop}$ as a function of $ | \cos[ \pi (\Phi_1/\Phi_0-\theta)] | $ where $\theta$ is a shift in $\Phi_1$. The aforementioned conditions take a more eloquent form when expressed in terms of the Josephson critical currents of each junction, giving 

%The three cases plotted in Fig. \ref{Fig3} satisfy the aforementioned conditions and shall therefore be useful to illustrate this behavior. In Fig. \ref{Fig4}(b) we plot $\dot{Q}_{int}(\Phi_1)$ for $\Phi_2=0$ (dashed lines) and for the corresponding value of $\Phi_2$ that provides the fully suppression of $\dot{Q}_{int}$ (solid lines). The curves corresponding to the symmetric double-loop have already been plotted in Fig. \ref{Fig2}(a) and \ref{Fig2}(b), and are therefore not shown. Notably, when setting $r_1=0.5$ and $r_2=0.5$, i.e., $R_a=R_c=2R_b$, $\dot{Q}_{int}$ cancels precisely when $\Phi_2= \Phi_0/2$ at $\Phi_1= \Phi_0/2 \pm k \Phi_0$ since this case satisfies exactly the condition $r_1 + r_2 = 1$. Notice that the total amplitude of the oscillation is reduced by one half with respect to the previous example. Let us finally consider a last illustrative case belonging to the region defined by $1 \leq r_1 \leq r_2+ 1$, i.e, the diagonally stripped region in Fig. \ref{Fig4}(a). If $r_1=1.25$ and $r_2=0.75$, i.e., $R_a=0.6R_c$ and $2R_b=0.75R_c$, although corresponding to a substantially asymmetric interferometer, it should  be possible to suppress completely $\dot{Q}_{int}$ while conserving the maximum amplitude of oscillation. As we can see in the bottom panel of Fig. \ref{Fig4}(b) this is exactly the case. Setting $\Phi_2= \Phi_0/4$, the phase-dependent component of heat transport cancels at $\Phi_1=3 \Phi_0 /5 \pm k \Phi_0$ with $\delta \dot{Q}_{int} = 2\dot{Q}_{int}^a$.

\subsection{Experimental realization proposal}
\label{subsec22}

Let us finally analyze how the previously described control over the magnetic flux-to-heat current transfer functions are translated into a realistic situation. For this purpose, a device similar to that of Fig. \ref{Fig13}(a) could be envisioned. In this specific case, the single SQUID loop must be substituted by a double-loop consisting of three parallel-connected Josephson junctions and  with independent magnetic flux controls. Similar devices operating as charge interferometers \cite{Kemppinen} have been reported already in the literature proving the feasibility of this structure. %Also in this case, a practical device can be easily fabricated by standard electron-beam lithography and shadow mask evaporation of superconducting metals, e.g., aluminum. Aluminum oxide for the Josephson barriers and copper for the normal metal source and drain electrodes can be used. A temperature gradient can easily be established across the superconducting double loop by intentionally heating one double-loop SQUID branch while maintaining the other well thermalized at the minimum bath temperature. Temperature detection and manipulation can be performed through two normal metal leads tunnel-coupled to one superconducting electrode of the double-loop SQUID that allow for the implementation of normal metal-insulator-superconductor thermometers and heaters \cite{RMP}.

%____________________________________________________________________________________________________________________________________________________________
\begin{figure}[t]
\centering
\includegraphics[width=1\columnwidth]{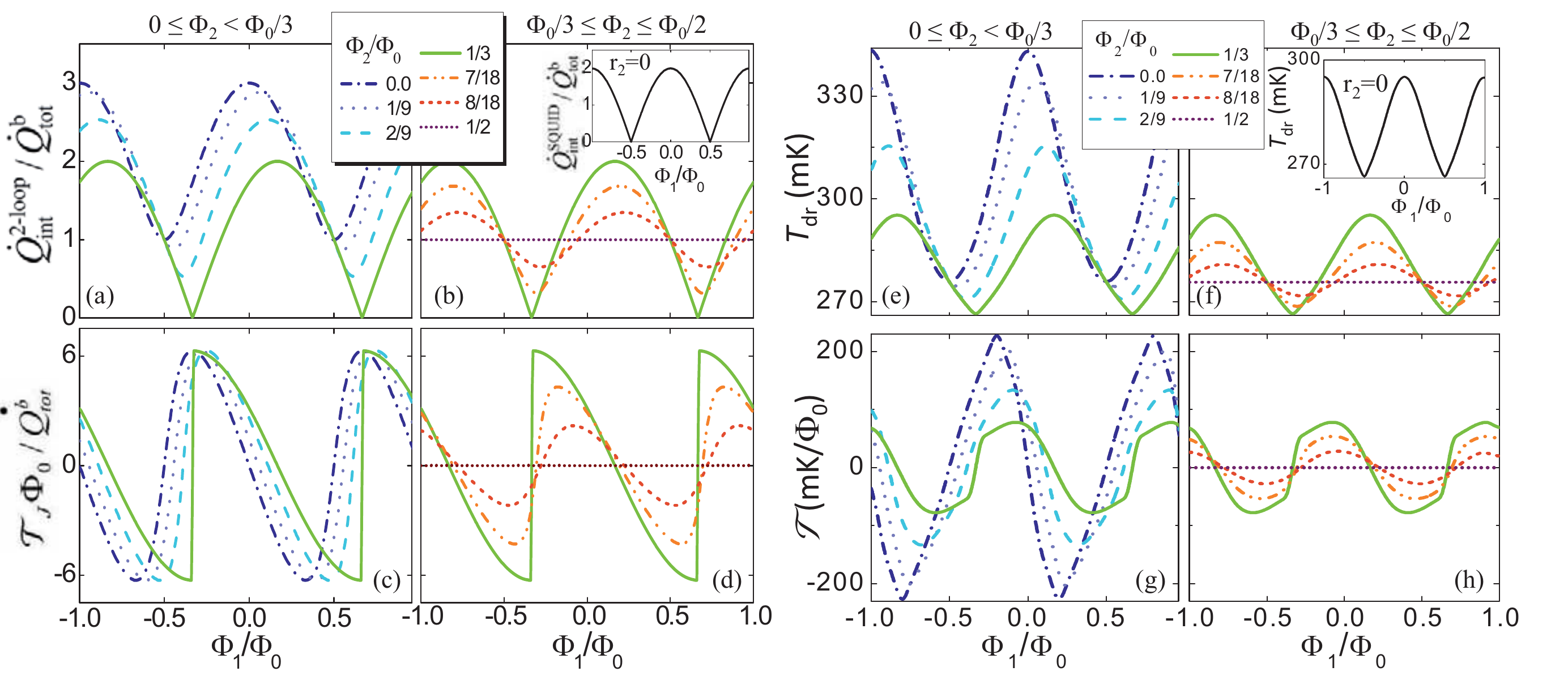}
\caption{(Color online) Panels (a) and (b): $\dot{Q}_{int}^{\rm 2-loop}$ vs. $\Phi_1$ plotted for different values of the control flux $\Phi_2$ for  $r_1=r_2=1$. The inset shows the case for $r_2=0$. Panels (c) and (d): transfer function $\mathcal{T} _{\rm J} = \partial \dot{Q}_{int}^{\rm 2-loop} / \partial \Phi_1 $ vs. $\Phi_1$ plotted for the same values of $\Phi_2$ as in the top panels. Panels (e) and (f): $T_{\rm dr}$ modulation calculated using the thermal model depicted in Fig. \ref{Fig21}(b) for the same conditions as in panels (a) and (b). Panels (e) and (f): Corresponding flux-to-temperature transfer coefficient.}
\label{Fig22}
\end{figure}
%____________________________________________________________________________________________________________________________________________________________

The device we envision is that schematized in Fig. \ref{Fig21}(a). As it was argued in section \ref{subsec12} heat transport in this structure can be studied by intentionally heating electrons in the source up to $T_{\rm src}$ yielding a quasiparticle temperature $T_{\rm hot}>T_{\rm cold}=T_{\rm bath}$ in S$_1$,  therefore leading to a finite heat current $\dot{Q}_{\rm 2-loop}$. The latter can be modulated by means of the two control magnetic fluxes $\Phi_1$ and $\Phi_2$ and inferred by measuring the drain electron temperature ($T_{\rm dr}$). 

Drain temperature can be predicted by solving a couple of thermal balance equations similar to those described in  section \ref{subsec12}. New equations will differ in that no additional probe [i.e., $S_3$ electrode in Fig. \ref{Fig13}(b)] is considered this time. For this reason it will be convenient to include now the electron-phonon interaction in $S_1$ [$\dot{Q}_{\rm e-ph, S_1}$, defined in Eq. (\ref{eph})]. %  , no additional probe [i.e., $S_3$ in Fig. \ref{Fig13}(b)] is considered. It will be convenient therefore to take into accounting  for the main heat exchange mechanisms existing in the structure, according to the model shown in Fig. \ref{Fig21}(b). Unlike the model proposed in section \ref{subsec12}, the new model does not include   S$_1$ exchanges heat with source electrons at power $\dot{Q}_{\rm src}$, with drain at power $\dot{Q}_{\rm dr}$, and with quasiparticles in S$_2$ at power $\dot{Q}_{\rm 2-loop}$.  Furthermore, electrons in the structure exchange heat with lattice phonons residing at bath temperature $T_{\rm bath}$, in particular, at power $\dot{Q}_{\rm e-ph, S_1}$ in S$_1$, and at power $\dot{Q}_{\rm e-ph,src}$  and $\dot{Q}_{\rm e-ph,dr}$ in source and drain electrodes, respectively.  Finally, we assume S$_2$ to be large enough to provide substantial electron-phonon coupling $\dot{Q}_{\rm e-ph, S_2}$ so that its quasiparticles will reside at $T_{\rm bath}$.
%Finally, we assume quasiparticles in S$_2$ to be thermalized at $T_{\rm bath}$ thanks to a large $\dot{Q}_{e-ph, S_2}$ heat current. 
In this way, we have now $-\dot{Q}_{\rm src}+\dot{Q}_{\rm 2-loop}+\dot{Q}_{\rm dr}+\dot{Q}_{\rm e-ph, S_1}=0$ and $-\dot{Q}_{\rm dr}+\dot{Q}_{\rm e-ph,dr}=0$
for S$_1$ and drain, respectively.\footnote{%The expression of $\dot{Q}_{\rm src}$, $\dot{Q}_{\rm src}$ and $\dot{Q}_{\rm e-ph,dr}$ are those given in section \ref{subsec12} whereas $\dot{Q}_{\rm 2-loop}$ is that given in Eq. \ref{q2loop}. 
As a set of parameters representative for a realistic microstructure we choose $R_{\rm src}=R_{\rm dr}=2$ k$\Omega$, $R_{\rm J}=500$ $\Omega$, $\mathcal{V}_{\rm dr}=10^{-20}$ m$^{-3}$, $\Sigma_{\rm dr}=3\times 10^9$ WK$^{-5}$m$^{-3}$ (typical of Cu) \cite{RMP}, $\mathcal{V}_{\rm S_1}=10^{-18}$ m$^{-3}$, $\Sigma_{\rm S_1}=3\times 10^8$ WK$^{-5}$m$^{-3}$ and $\Delta_1(0)=\Delta_2(0)=200\,\mu$eV, the last two parameters typical of aluminum (Al) \cite{RMP}.}

The results of thermal balance equations for drain temperature are shown in Fig. \ref{Fig22}(e) and (f) for  $r_1=r_2=1$ and the same $\Phi_2$ values as in panels (a) and (b). For these calculations we have set $T_{\rm bath}=245$ mK and $T_{\rm src}=700$ mK. As it can be seen, $T_{\rm dr}$ is $\Phi_1$ periodic  reaching its minimum value for $\Phi_2 = \Phi_0 /3$, as expected, but the appearance of the oscillation depends strongly on $\Phi_2$. We highlight that, for $\Phi_2 = 0$, the $T_{\rm dr}$ modulation is, under equal conditions, more than twice that obtained with the single thermal interferometer discussed in the previous section [see inset in Fig. \ref{Fig22}(f)]. This enhancement is related with the overall enhancement of $\dot{Q}_{int}^{\rm 2-loop}$ and the fact that  $T_{\rm dr}$ does not depend linearly on  $\dot{Q}_{\rm 2-loop}$. In this way, by simultaneously playing with both $\Phi_1$ and $\Phi_2$, $T_{\rm dr}$ could be modulated from $\sim 265$ up to $\sim 345$ mK in overall. Differences between the modulation curves obtained at each  $\Phi_2$ values are emphasized in Fig. \ref{Fig22}(g) and (h) where the flux-to-temperature transfer coefficient $\mathcal T$ is plotted.

\section{Heat diffractor}
\label{sec4}

In this section we theoretically analyze heat transport in temperature-biased extended JJ showing that the phase-dependent component of thermal flux  through the weak-link interferes in the presence of an in-plane magnetic field  leading to \emph{heat diffraction} \cite{solinas}, in analogy to what occurs for the Josephson \textit{electric} critical current \cite{rowell}.
In particular, thermal transport is investigated in three prototypical \emph{electrically-open} junctions geometries showing that the quantum phase difference across the junction undergoes $\pi$ \emph{jumps} in order to minimize the Josephson coupling energy.
We finally  propose how to demonstrate thermal diffraction and to prove the existence of $\pi$ jumps  in a realistic microstructure.

%%%%%%%%%%%%%%%
%%%%%%%%%%%%%%%%%%FIGURE 1%%%%%%%%%%%%%%%%%%
%\begin{figure}[t!]
%\includegraphics[width=\columnwidth]{fig1.pdf}
%\vspace{-4mm}
%\caption{(Color online) (a) Cross section of a temperature-biased extended  S$_1$IS$_2$ Josephson tunnel junction in the presence of an in-plane magnetic field $H$. 
%The heat current $\dot{Q}_{S_1\rightarrow S_2}$ flows along the $z$ direction whereas $H$ is applied in the $x$ direction, i.e., parallel to a symmetry axis of the junction. 
%Dashed line indicates the closed integration contour, $T_k$, $t_k$ and $\lambda_k$ represent  the temperature, thickness and London penetration depth of superconductor S$_k$, respectively, and $d$ is the insulator thickness. $\Phi$ denotes the magnetic flux piercing the junction.  
%Prototypical junctions with rectangular, circular, and annular geometry are shown in panel (b), (c) and (d), respectively. $L$, $W$, $R$ and $r$ represent the junctions geometrical parameters. 
%}
%\label{fig1}
%\end{figure}
%%%%%%%%%%%%%%%
%%%%%%%%%%%%%%%%%%%%
\subsection{Theoretical considerations and layout design}
\label{model}
\begin{figure}[t]
\centering
\includegraphics[width=0.9\columnwidth]{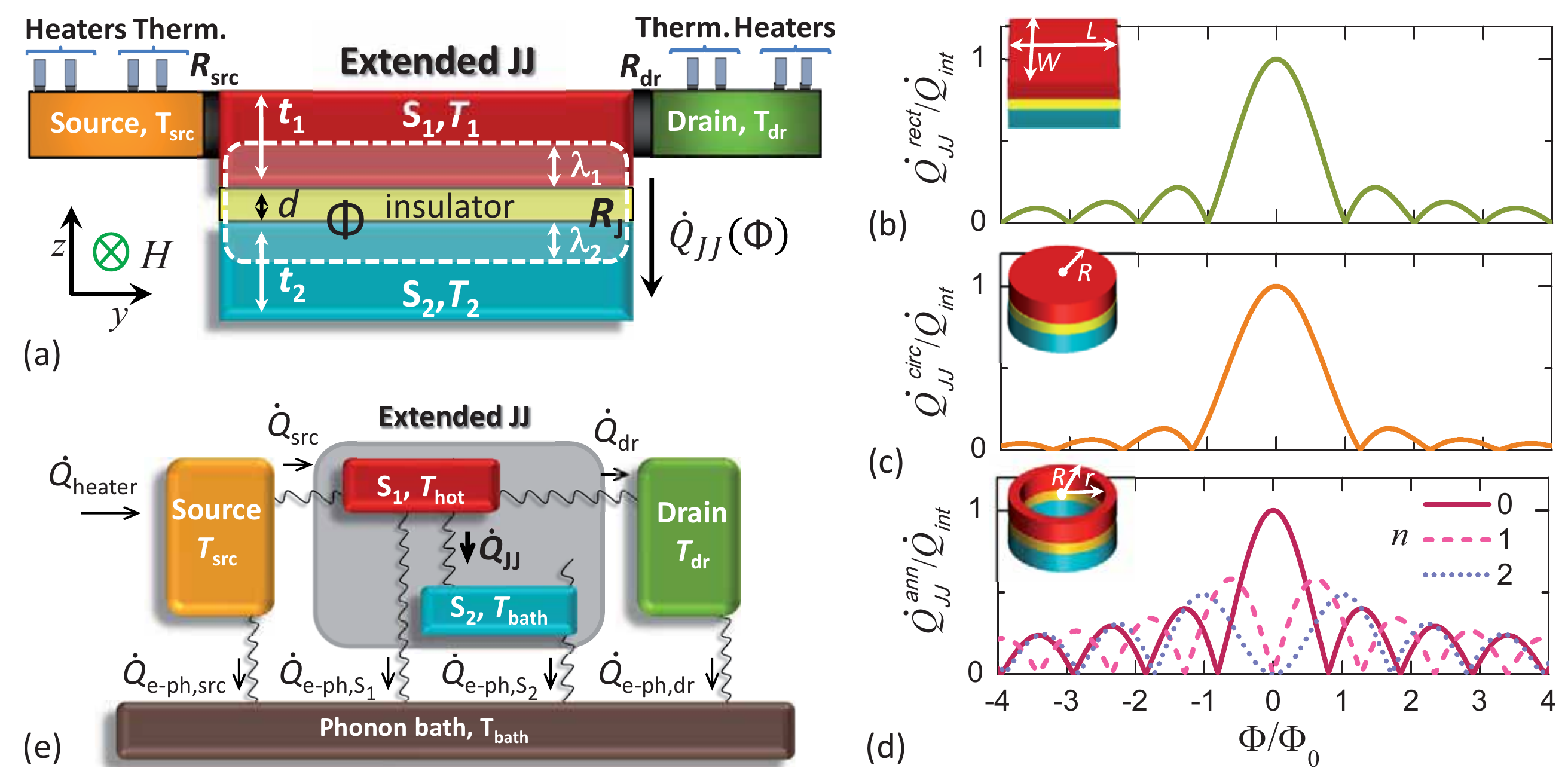}
\caption{(Color online) (a) Cross section of a temperature-biased extended  JJ  parallel to the $xy$ plane in the presence of an in-plane magnetic field $H$ along the $x$ direction. Additionally, source and drain normal-metal electrodes are tunnel-coupled to one of the junction electrodes (S$_1$). Superconducting tunnel junctions operated as heaters and thermometers are connected to source and drain.   
$\dot{Q}^{\rm JJ}_{int}$vs. $\Phi$ calculated for a rectangular [(a)], circular [(b)], and annular [(c)] JJ represented as insets. $L$, $W$, $R$ and $r$ represent the junctions geometrical parameters.  
(e) Thermal model describing the main heat exchange mechanisms existing in the structure shown in (a).}
\label{Fig31}
\end{figure}

Our system is schematized in Fig. \ref{Fig31}(a), and consists of an extended JJ composed of two superconducting electrodes S$_1$ and S$_2$ residing at different temperatures $T_{\rm hot}$ and $T_{\rm cold}$, respectively. S$_1$ and S$_2$ are characterized by their London penetration depths $\lambda_1$ and $\lambda_2$, respectively, and separated by an insulating layer with thickness $d$.  An external magnetic field ($H$) is applied parallel to a symmetry axis of the junction [see Fig. \ref{Fig31}(a)] penetrating within a distance $t_H=\lambda_1+\lambda_2+d$ in the $z$ direction\footnote{When this condition is no longer satisfied an effective thickness ($\tilde{t}_H$) must be used, where $\tilde{t}_H=\lambda_1\textrm{tanh}(T_{\rm hot}/2\lambda_1)+\lambda_2\textrm{tanh}(T_{\rm hot}/2\lambda_2)+d$ \cite{Weihnacht1969}.}.
We will consider symmetric JJs in the \emph{short} limit and neglect the effect of the edges.

%____________________________________________________________________________________________________________________________________________________________
%____________________________________________________________________________________________________________________________________________________________

%The total heat current flowing through the extended JJ reads
%\begin{equation}
%\dot{Q}_{\rm JJ}=\dot{Q}_{qp}^{\rm JJ}(T_{\rm hot},T_{\rm cold})-\dot{Q}^{\rm JJ}_{int}(T_{\rm hot},T_{\rm cold},H).
%\label{qdot}
%\end{equation}

We shall focus onto the phase-dependent component only. 
To this end we need to determine the phase gradient $\varphi (x,y)$ induced by the application of $H$. 
By choosing the closed integration contour indicated by the dashed line depicted in Fig. \ref{Fig31}(a)
it can be shown \cite{Tinkham,Barone} that, neglecting screening induced by the Josephson current, %$\varphi (x,y)$ obeys the equations $\partial \varphi/\partial x=0$ and $\partial \varphi/\partial y=2\pi \mu_0 t_H H/\Phi_0$. The latter equation can be easily integrated to yield  
$\varphi (y)=\kappa y+\varphi_0$,
where $\kappa \equiv 2\pi \mu_0 t_H H/\Phi_0$ and $\varphi_0$ is the phase difference at $y=0$. 
After integration over $x$ the phase-dependent component of the heat current can  then be written as $\dot{Q}^{\rm JJ}_{int}(T_{\rm hot},T_{\rm cold},H)=\textrm{Re}\big\{e^{i\varphi_0}\int^{\infty}_{-\infty} dy \mathcal{Q}(y,T_{\rm hot},T_{\rm cold})e^{i\kappa y}\big\}$,
%\begin{equation}
%\dot{Q}^{\rm JJ}_{int}(T_{\rm hot},T_{\rm cold},H)=\int\int dxdy \dot{Q}_A(x,y,T_{\rm hot},T_{\rm cold})\textrm{cos}(\kappa y+\varphi_0),
%\label{phasea}
%\end{equation}
%where the integration is performed over the junction area, and $\dot{Q}_A(x,y,T_{\rm hot},T_{\rm cold})$ is the heat current density per unit area. 
%We note that the integrand of Eq. (\ref{phasea}) oscillates sinusoidally along the $y$ direction with period given by $\Phi_0(\mu_0t_H H)^{-1}$.
%After integration over $x$ we can write Eq. (\ref{phasea}) as 
%\begin{eqnarray}
%\dot{Q}^{\rm JJ}_{int}(T_{\rm hot},T_{\rm cold},H)=\textrm{Re}\left\{e^{i\varphi_0}\int^{\infty}_{-\infty} dy \mathcal{Q}(y,T_{\rm hot},T_{\rm cold})e^{i\kappa y}\right\},
%\label{phaseb}
%\end{eqnarray}  
%&=&\int dy \mathcal{Q}(y,T_{\rm hot},T_{\rm cold})\textrm{cos}(\kappa y+\varphi_0)\nonumber\\
where $\mathcal{Q}(y,T_{\rm hot},T_{\rm cold})$%$\equiv \int dx \dot{Q}_A(x,y,T_{\rm hot},T_{\rm cold})$ 
 is the heat current density integrated along $x$. 
%In writing  second equality in Eq. (\ref{phaseb}) we have replaced the integration limits by $\pm \infty$ since the thermal current is zero outside the junction.
This equation  resembles the expression for  the Josephson current given by $i_{\rm J}(T_{\rm hot},T_{\rm cold},H)=\textrm{Im}\big\{e^{i\varphi_0} \int^{\infty}_{-\infty} dy \mathcal{I}(y,T_{\rm hot},T_{\rm cold})e^{i\kappa y}\big\}$,
%\begin{eqnarray} 
%i_{\rm J}(T_{\rm hot},T_{\rm cold},H)=\textrm{Im}\left\{e^{i\varphi_0}\int^{\infty}_{-\infty} dy \mathcal{I}(y,T_{\rm hot},T_{\rm cold})e^{i\kappa y}\right\},%\textrm{sin}\varphi_0\int^{\infty}_{-\infty} dy\mathcal{I}(y)\textrm{cos}\kappa y,
%\label{critical}
%\end{eqnarray}
%&=&\textrm{Im}\left\{e^{i\varphi_0}\int^{\infty}_{-\infty} dy \mathcal{I}(y,T_{\rm hot},T_{\rm cold})e^{i\kappa y}\right\} \nonumber\\
 where $\mathcal{I}(y,T_{\rm hot},T_{\rm cold})$ is the supercurrent density integrated along $x$ \cite{Tinkham,Barone}.
%If $\mathcal{I}_s(y)$ and $\mathcal{I}_a(y)$ denote the symmetric and antisymmetric part of $\mathcal{I}(y)$, respectively, it follows that Eq. (\ref{critical}) can be re-written as 
%\begin{eqnarray}
%I_{\rm JJ}&=&\int^{\infty}_{-\infty} dy[\textrm{cos}\varphi_0\mathcal{I}_a(y)\textrm{sin}\kappa y+\textrm{sin}\varphi_0\mathcal{I}_s(y)\textrm{cos}\kappa y]\nonumber\\
%&=&\textrm{sin}\varphi_0\int^{\infty}_{-\infty} dy\mathcal{I}_s(y)\textrm{cos}\kappa y,
%\label{critical2}
%\end{eqnarray}
%and second equality in Eq. (\ref{critical}) follows from the assumed junctions \emph{symmetry}, i.e., $\mathcal{I}(y,T_{\rm hot},T_{\rm cold})=\mathcal{I}(-y,T_{\rm hot},T_{\rm cold})$.
In the actual configuration of electrically-open junction, the condition of \emph{zero} Josephson current for any given value of $H$
yields the solution $\varphi_0=m\pi$, with $m=0,\pm 1,\pm2\ldots$.  

On the other hand, the Josephson coupling energy of the junction ($E_{\rm J}$) can be expressed as $E_{\rm J}(T_{\rm hot},T_{\rm cold},H)=E_{J,0}-\frac{\Phi_0}{2\pi}\textrm{cos}\varphi_0\int_{-\infty}^{\infty} dy\mathcal{I}(y)\textrm{cos}\kappa y$
%\begin{eqnarray}
%E_{\rm J}(T_{\rm hot},T_{\rm cold},H)
%&=&E_{J,0}-\frac{\Phi_0}{2\pi}\int_{-\infty}^{\infty} dy[\textrm{cos}\varphi_0\mathcal{I}_s(y)\textrm{cos}\kappa y-\textrm{sin}\varphi_0\mathcal{I}_a(y)\textrm{sin}\kappa y]\nonumber\\
%=E_{J,0}-\frac{\Phi_0}{2\pi}\textrm{cos}\varphi_0\int_{-\infty}^{\infty} dy\mathcal{I}(y)\textrm{cos}\kappa y,
%\label{energy}
%\end{eqnarray}
%&=&E_{J,0}-\frac{\Phi_0}{2\pi}\textrm{Re}\left\{e^{i\varphi_0}\int_{-\infty}^{\infty} dy \mathcal{I}(y,T_{\rm hot},T_{\rm cold})e^{i\kappa y}\right\} \nonumber\\ 
where
$E_{J,0}=\Phi_0i_c/2\pi$, $i_c$ is the zero-field critical supercurrent. 
%and in writing the second equality we have used the symmetry property of $\mathcal{I}(y,T_{\rm hot},T_{\rm cold})$. 
Minimization of $E_{\rm J}$ for any applied $H$ imposes the second term of $E_{J,0}$ to be always negative, so that
 $\varphi_0$ will undergo a $\pi$ \emph{jump} whenever the integral does contribute to $E_{\rm J}$ with negative sign. 
%As a result, the Josephson  coupling energy turns out to be written as
%\begin{equation}
%E_{\rm J}(T_{\rm hot},T_{\rm cold},H)=E_{J,0}-\frac{\Phi_0}{2\pi}\left|\int_{-\infty}^{\infty} dy\mathcal{I}(y,T_{\rm hot},T_{\rm cold})\textrm{cos}\kappa y\right|. 
%\label{energybis}
%\end{equation} 
It can be shown that the $\pi$ jumps are present in any junction geometry \cite{solinas}.
They have energetic origin only. This fact makes them very different from those present in low dimensional superconductors,
caused by thermal \cite{Langer1967} and quantum \cite{Zaikin1997,astafiev2012} fluctuations.

Taking into account the aforementioned $\pi$ \emph{jumps} and assuming that the symmetry of the junction %and of the electric current  density  are 
is reflected in an analogous symmetry in the heat current, i.e., $\mathcal{Q}(y,T_{\rm hot},T_{\rm cold}) = \mathcal{Q}(-y,T_{\rm hot},T_{\rm cold})$, the phase-dependent component of the heat current can be written as $\dot{Q}^{\rm JJ}_{int}=\left|\int^{\infty}_{-\infty} dy \mathcal{Q}(y,T_{\rm hot},T_{\rm cold})\textrm{cos}\kappa y\right|$.

We can now determine the behavior of $\dot{Q}^{\rm JJ}_{int}$ for the three prototypical junction geometries sketched in Fig. (\ref{Fig31}). 
In particular, we shall consider two well-known examples such as the \emph{rectangular} [see Fig. \ref{Fig31}(b)] and \emph{circular} [see Fig. \ref{Fig31}(c)] junction, and the more exotic \emph{annular} one [see Fig. \ref{Fig31}(d)]. Annular junctions offer the possibility to investigate fluxons dynamics due to the absence of collisions with boundaries; yet, they provide fluxoid quantization thanks to their geometry, which allows fluxons trapping. 

We assume that the total phase-dependent heat current is characterized by a uniform distribution.
$\dot{Q}^{\rm JJ}_{int}$ can therefore be calculated for the three considered geometries by following, for instance, Refs. \cite{Tinkham,Barone}. 
In particular, for the rectangular junction,  the absolute value of the sine cardinal function is obtained, $\dot{Q}_{int}^{rect}(T_{\rm hot},T_{\rm cold},\Phi)=\dot{Q}_{int}(T_{\rm hot},T_{\rm cold})\left| \sin(\pi\Phi/\Phi_0)(\pi\Phi/\Phi_0)\right|$.
%\begin{equation}
%\dot{Q}_{int}^{rect}(T_{\rm hot},T_{\rm cold},\Phi)=\dot{Q}_{int}(T_{\rm hot},T_{\rm cold})\left|\frac{\textrm{sin}(\pi\Phi/\Phi_0)}{(\pi\Phi/\Phi_0)}\right|, 
%\end{equation}
%where $\dot{Q}_{int}(T_{\rm hot},T_{\rm cold})=WL\dot{Q}_A(T_{\rm hot},T_{\rm cold})$, $\Phi=\mu_0 HLt_H$, $L$ is the junction length and $W$ its width.
For the circular geometry one gets the Airy diffraction pattern, 
$\dot{Q}_{int}^{circ}=\dot{Q}_{int}\left|\dot{Q}_1(\pi\Phi/\Phi_0)/(\pi\Phi/2\Phi_0)\right|$,
where %$\dot{Q}_{int}(T_{\rm hot},T_{\rm cold})=\pi R^2 \dot{Q}_A(T_{\rm hot},T_{\rm cold})$, 
$\dot{Q}_1(y)$ is the Bessel function of the first kind, $\Phi=2\mu_0 HRt_H$, and $R$ is the junction radius.
Finally, for the annular junction \cite{Martucciello1996,Nappi1997}, the phase-dependent component of the heat current takes the form
$\dot{Q}_{int}^{ann}=2\dot{Q}_{int}/(1-\alpha^2)\left|\int^1_\alpha dxx\dot{Q}_n(x\pi \Phi/\Phi_0)\right|$, 
where %$\dot{Q}_{int}(T_{\rm hot},T_{\rm cold})=\pi(R^2-r^2)\dot{Q}_A(T_{\rm hot},T_{\rm cold})$, 
$\Phi=2\mu_0 HRt_H$,
$\alpha=r/R$, $\dot{Q}_n(y)$ is the $n$th Bessel function of integer order, $R$ ($r$) is the external (internal)  radius, and $n=0,1,2,...$ is the number of $n$ trapped fluxons in the junction barrier.

%____________________________________________________________________________________________________________________________________________________________
%____________________________________________________________________________________________________________________________________________________________
%\begin{figure}[t]
%\centering
%\includegraphics[width=0.9\columnwidth]{Fig32}
%\caption{(Color online) Normalized phase-dependent  component of the heat current $\dot{Q}^{\rm JJ}_{int}$ versus magnetic flux $\Phi$ calculated for a rectangular [(a)], circular [(b)], and annular [(c)] JJ. Results from the thermal model shown in Fig. \ref{Fig31}(f) showing the $T_{\rm dr}$ modulation for  a rectangular [(d)], circular [(e)], and annular [(f)] JJ. For this calculations we set $T_{\rm bath}=245 $ mK and $T_{\rm src}=700 $ mK. In the curves of panel (c) and (f) we set $\alpha=0.9$, and $n$ indicates the number of fluxons trapped in the junction barrier.  }
%\label{Fig32}
%\end{figure}
%____________________________________________________________________________________________________________________________________________________________
%____________________________________________________________________________________________________________________________________________________________

Right panels in Fig. \ref{Fig31} illustrate the behavior of $\dot{Q}^{\rm JJ}_{int}$ for the three geometries. In particular, the curve displayed in Fig. \ref{Fig31}(a) for the rectangular case shows the well-known Fraunhofer diffraction pattern analogous to that produced by light diffraction through a rectangular slit exhibiting minima for integer multiples of $\Phi_0$. %Furthermore, the heat current is  rapidly damped by increasing the magnetic field falling asymptotically as $\Phi^{-1}$ \cite{Tinkham}.
In the case of a circular junction, the flux values where $\dot{Q}^{circ}_{int}$ vanishes do not coincide anymore with multiples of $\Phi_0$, and $\dot{Q}^{circ}_{int}$ falls more rapidly than in the rectangular junction case. %, i.e., as $\Phi^{-3/2}$ \cite{Tinkham}.
Finally, for an annular junction, the heat current diffraction pattern is strongly $n$-dependent providing, apparently, enhanced flexibility to tailor the  heat current response.
%, differently from the rectangular and circular case, $\dot{Q}^{ann}_{int}$ decays in general more slowly. It is apparent that annular junctions may provide, in principle, enhanced flexibility to tailor the  heat current response. 
 
\subsection{Proposed experimental setup}
\label{experiment}

Demonstration of diffraction of thermal currents could be achieved in the setup shown in Fig. \ref{Fig31}(a). This is similar to the device already analyzed in section \ref{subsec22} but the in-plane static magnetic field $H$ is applied perpendicular to the Josephson weak-link. The experimental operation mode will be the same and, again, the drain temperature behavior can be predicted by solving a couple of thermal balance equations accounting  for the main heat exchange mechanisms existing in the structure, according to the model shown in Fig. \ref{Fig31}(e).\footnote{Calculations are performed using the same fabrication parameters of \ref{subsec22}, i.e., $R_{\rm src}=R_{\rm dr}=2$ k$\Omega$, $R_{\rm J}=500$ $\Omega$, $\mathcal{V}_{dr}=10^{-20}$ m$^{-3}$, $\Sigma_{dr}=3\times 10^9$ WK$^{-5}$m$^{-3}$, $\mathcal{V}_{S_1}=10^{-18}$ m$^{-3}$, $\Sigma_{S_1}=3\times 10^8$ WK$^{-5}$m$^{-3}$ and $\Delta_1(0)=\Delta_2(0)=200\,\mu$eV.}

%_______________________________________________________________________________________________________________________________________________________
%_______________________________________________________________________________________________________________________________________________________%$\begin{figure}[t]
\begin{figure}[t]
\centering
\includegraphics[width=1\columnwidth]{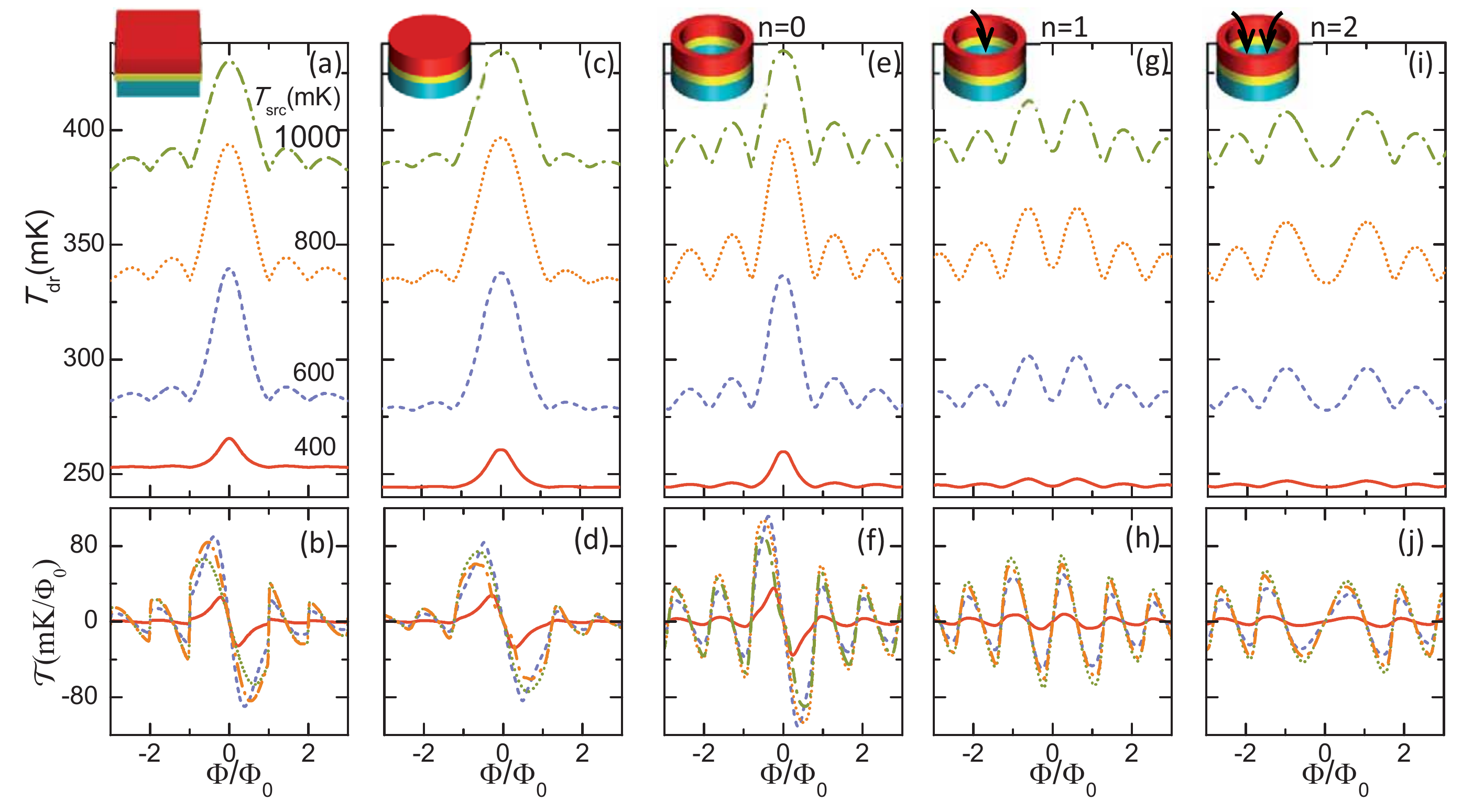}
\caption{(Color online) (a) $T_{dr}$ vs $\Phi$ calculated at $T_{\rm bath}=245$ mK for several values of source temperature $T_{\rm src}$ for a structure based on a a rectangular [(a)], circular [(c)], and annular with $n=0$ [(e)], $n=1$ [(g)] and $n=2$ [(i)]. Bottom panels show the corresponding flux-to temperature transfer function $\mathcal{T}$ vs $\Phi$. For the annular case we set $\alpha=0.9$. }
\label{Fig33}
\end{figure}
%______________________________________________________________________________________________________________________________________________________
%______________________________________________________________________________________________________________________________________________________

The results of thermal balance equations for drain temperature and different source temperatures are shown in Fig. \ref{Fig33}. As expected, $T_{\rm dr}$ shows a response to magnetic flux clearly resembling a Fraunhofer or an Airy-like diffraction patterns for the rectangular and circular cases, respectively. For the annular case the temperature patterns are drastically different depending on the number of fluxons trapped inside the loop. In all cases, temperature minima appear at the expected  $\Phi_0$ positions being the the unequivocal manifestation of the above-described phase \emph{jumps}. 

%The impact of an increasing source temperature on the structure response is shown in the top panels of Fig. \ref{Fig33} where $T_{\rm dr}$ is plotted against $\Phi$ for a few $T_{\rm src}$ values. Increasing $T_{\rm src}$ leads to a monotonic enhancement of the maximum of $T_{\rm dr}$  at $\Phi=0$ which stems from an increased heat current flowing into drain electrode. Furthermore, the amplitude of $T_{\rm dr}$ lobes follows a non-monotonic beahavior, initially increasing with source temperature, being maximized at intermediate temperatures, and finally decreasing at higher $T_{\rm src}$ values. 

The flux-to-temperature transfer coefficient, $\mathcal{T}=\partial T_{\rm dr}/\partial \Phi$, corresponding to each of the above-mentioned cases is plotted in  the bottom panels of Fig. \ref{Fig33}. As it can be seen there, $\mathcal{T}$ exceeding $\sim 100$ mK$/\Phi_0$ could be achieved at $T_{\rm src}=600$ mK for the annular case. Moreover, the transfer coefficient clearly demonstrates the non-monotonicity of the amplitude of drain temperature lobes as a function of $T_{\rm src}$.

%%%%%%%%%%%%%%%%%%%%%%%%%%%%%%%%%%%%%%%%%%%%%%%%%%%%%%%%%%%%%%%%%%%%%%%%%%%%%%%%%%%%%%%%%%%%%%%%%%%%%%%%%%%%%%%%%%%%%%%%%%%%%%%%%%%%%%%%%%%%%%%%%%%%%%%%%%%%%%%%%%%%%%%%%%%%%%%%%%%%%%%%%%%%%%%%%%%%%%%%%%%%%%%%%%%%%%%%%%%%%%%%%%%%%%%%%%%%%%%%%%%%%%%%%%%%%%%%%%%%%%%%%%%%%%%%%%%%%%%%%%%%%%%%%%%%%%%%%%%%%%%%%%%%%%%%%%%%%%%%%%%%%%%%%%%%%%%%%%%%%%%%%%%%%%%%%%%%%%%%%%%%%%%%%%%%%%%%%%%%%%%%%%%%%%%%%%%%%%%%%%%%%%%%%%%%%%%%%%%%%%%%%%%%%%%%%%%%%%%%%%%%%%%%%%%%%%%%%%%%%%%%%%%%%%%%%%%%%%%%%%%%%%%%%%%%%%%%%%%%%%%%%%%%%%%%%%%%%%%%%%%%%%%%%%%%%%%%%%%%%%%%%%%%%%%%%%%%%%%%%%%%%%%%%%%%%%%%%%%%%%%%%%%%%%%%%%%%%%%%%%%%%%%%%%%%%%%%%%%%%%%%%%%%%%%%%%%%%%%%%%%%%%%%%%%%%%%%%%%%%%%%%%%%%%%%%%%%%%%%%%%%%%%%%%%%%%%%%%%%%%%%%%%%%%%%%%%%%%%%%%%%%%%%%%%%%%%%%%%%%%%%%%%%%%%%%%%%%%%%%%%%%%%%%%%%%%%%%%%%%%%%%%%%%%%%%%%%%%%%%%%%%%%%%%%%%%%%%%%%%%%%%%%%%%%%%%%%%%%%%%%%%%%%%%%%%%%%%%%%%%%%%%%%%%%%%%%%%%%%%%%%%%%%%%%%%%%%%%%%%%%%%%%%%%%%%%%%%%%%%%%%%%%%%%%%%%%%%%%%%%%%%%%%%%%%%%%%%%%%%%%%%%%%%%%%%%%%%%%%%%%%%%%%%%%%%%%%%%%%%%%%%%%%%%%%%%%%%%%%%%%%%%%%%%%%%%%%%%%%%%%%%%%%%%%%%%%%%%%%%%%%%%%%%%%%%%%%%%

%\subsection{Conclusions}
%\label{concyu}

\section{Heat diode}
\label{sec5}

%So far, a strong effort has been devoted to envision thermal rectifiers, i.e., structures allowing high heat conduction along one direction but suppressed thermal transport upon temperature bias reversal \cite{RobertsRev,CasatiNandV}. Most of these proposals deal with phononic heat transport \cite{Li,Segal,Segal3,Terraneo}, very few deal with electronic heat conduction \cite{Ruokola1,Kuo,Ruokola,Chen} and even less have demonstrated feasible experimental realizations \cite{Scheibner,Chang}.   

In this section we propose and analyze theoretically the performance of two different thermal diodes. One, consisting of a NIS junction, and a second one, consisting of a SIS' Josephson tunnel junction \cite{Martinez2013}. Although never considered in the literature so far for such a purpose \cite{RobertsRev}, superconducting tunnel junctions appear particularly well suited for the implementation of electron heat rectifiers. Heat transport in such structures is deeply influenced by the strong temperature dependence of the superconducting density of states. %Remarkably large heat rectification coefficients up to $\sim 800\%$ can be theoretically achieved with a SIS'-based heat diode  using conventional materials and standard fabrication methods. 
In addition, the SIS' diode allows for the \emph{in-situ} fine tuning of the thermal rectification magnitude and direction.

%____________________________________________________________________________________________________________________________________________________________
%____________________________________________________________________________________________________________________________________________________________
\begin{figure}[t]
\centering
\includegraphics[width=0.8\columnwidth]{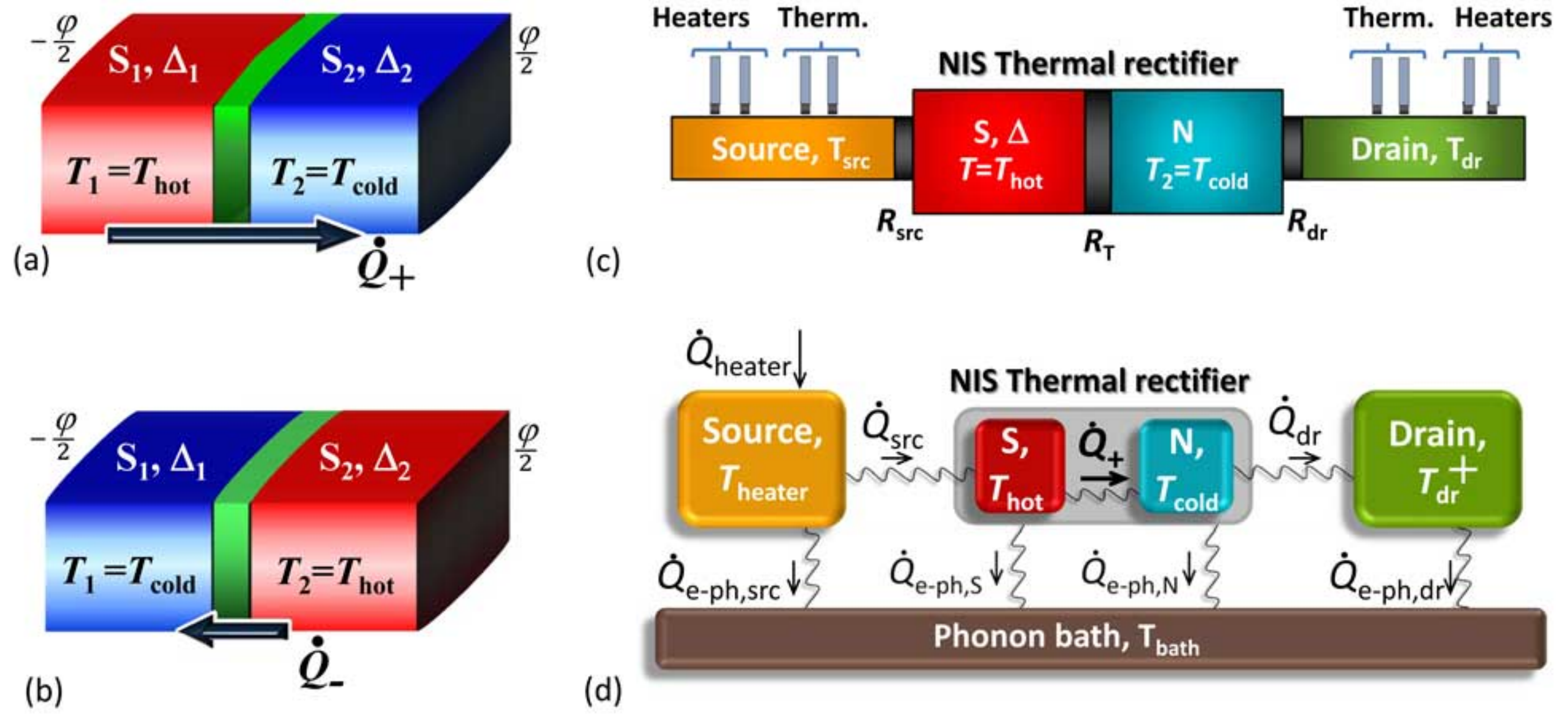}
\caption{(Color online) (a) and (b) Josephson thermal diode scheme corresponding to the forward and reverse thermal bias configuration, respectively. %The diode is phase biased ($\varphi$) and temperature biased with $T_{\rm hot} \neq T_{\rm hot}$  as well, but the voltage across the junction vanishes.  
(c) Possible experimental realization of the NIS thermal rectifier. Source and drain normal metal electrodes are tunnel-coupled to the core of the diode. Additional superconducting probes tunnel-coupled to source and drain allow for the implementation of SINIS thermometers and heaters. (d) Idealized thermal model including the main heat exchange mechanisms existing in the structure schematized in ( c).}
\label{Fig41}
\end{figure} 
%____________________________________________________________________________________________________________________________________________________________
%____________________________________________________________________________________________________________________________________________________________

We shall start, first of all, by defining a heat rectification parameter $\mathcal{R}$. To this end, let us consider a tunnel JJ made by two different superconductors, S$_1$ and S$_2$ characterized by its energy gap $\Delta_1$  and $\Delta_2$ leading to critical temperatures $T_{c_1}$ and $T_{c_2}$, respectively. The electronic temperature in both S$_1$ and S$_2$ is kept at fixed  $T_{\rm hot}$  and $T_{\rm cold}$, respectively, and the voltage drop across the junction is set to zero.  Additionally, $\varphi$ denotes the macroscopic phase difference across the junction with normal-state resistance $R_{\rm J}$. In the forward thermal bias configuration, a thermal gradient is created by setting $T_{\rm 1} = T_{\rm hot}  > T_{\rm 2}= T_{\rm cold}$, which leads to a total heat current $\dot{Q}_{+}$ flowing from S$_1$ to S$_2$ [see Fig. \ref{Fig41}(a)]. In the reverse thermal bias configuration, the thermal gradient is inverted so that $T_{\rm 1}= T_{\rm cold}< T_{\rm 2}= T_{\rm hot}$  leading to a heat current  $\dot{Q}_{-}$ flowing from S$_2$ to S$_1$ [see Fig. \ref{Fig41}(b)]. Under these hypothesis we define the rectification coefficient as $\mathcal{R} (\%)= (\dot{Q}_{+}-\dot{Q}_{-})\times100/\dot{Q}_{-}$. 
%Notice that $\dot{Q}_{-}$ is negative when referred  to S$_1$ (meaning that it is entering into S$_1$) so that $\mathcal{R}$ expresses  how larger is the forward heat current with respect to the reverse one. 
%For an ideal thermal rectifier, $\mathcal{R}$ may approach infinity as $\dot{Q}_{+} >> \dot{Q}_{-}$. %On the other hand, if  $\dot{Q}_{-}$ approaches $\dot{Q}_{+}$ the device does not rectifie giving   $\mathcal{R} =0$. Finally, 
%If the rectification direction switches $\mathcal{R}$ drops to a minimum value of $-100$ $\%$ as $\dot{Q}_{-} >> \dot{Q}_{+}$.

%In this way, $\mathcal{R}$ expresses  how larger is the forward heat current with respect to the reverse one. For an ideal thermal rectifier, $\mathcal{R}$ may approach infinity as $\dot{Q}_{+} \gg \dot{Q}_{-}$. On the other hand, if  $\dot{Q}_{-}$ approaches $\dot{Q}_{+}$ the device does not rectify giving   $\mathcal{R} =0$. Finally, If the rectification direction switches $\mathcal{R}$ drops to a minimum value of $-100$ $\%$ as $\dot{Q}_{-} \gg \dot{Q}_{+}$.

\subsection{NIS thermal diode}
\label{subsectionNIS}

We start by analyzing the case in which one of the two electrodes is a normal metal, i.e., a NIS junction. For this purpose we can simply set $\Delta_2 = 0$, which leads to the complete suppression of the interference component of the heat current, $\dot{Q}_{int}$. This leads to 
$  \dot{Q}_{+}=\frac{2}{e^2 R_{\rm T}}\int^{\infty}_{0} d\varepsilon \varepsilon \mathcal{N}_1 (\varepsilon,T_{\rm hot})[f(\varepsilon,T_{\rm cold})-f(\varepsilon,T_{\rm hot})]$ and 
$ \dot{Q}_{-}=\frac{2}{e^2 R_{\rm T}}\int^{\infty}_{0} d\varepsilon \varepsilon \mathcal{N}_1 (\varepsilon,T_{\rm cold})[f(\varepsilon,T_{\rm cold})-f(\varepsilon,T_{\rm hot})]$,  
where $R_{\rm T}$ is the tunnel resistance of the NIS junction.

\begin{figure}[t]
\centering
\includegraphics[width=0.85\columnwidth]{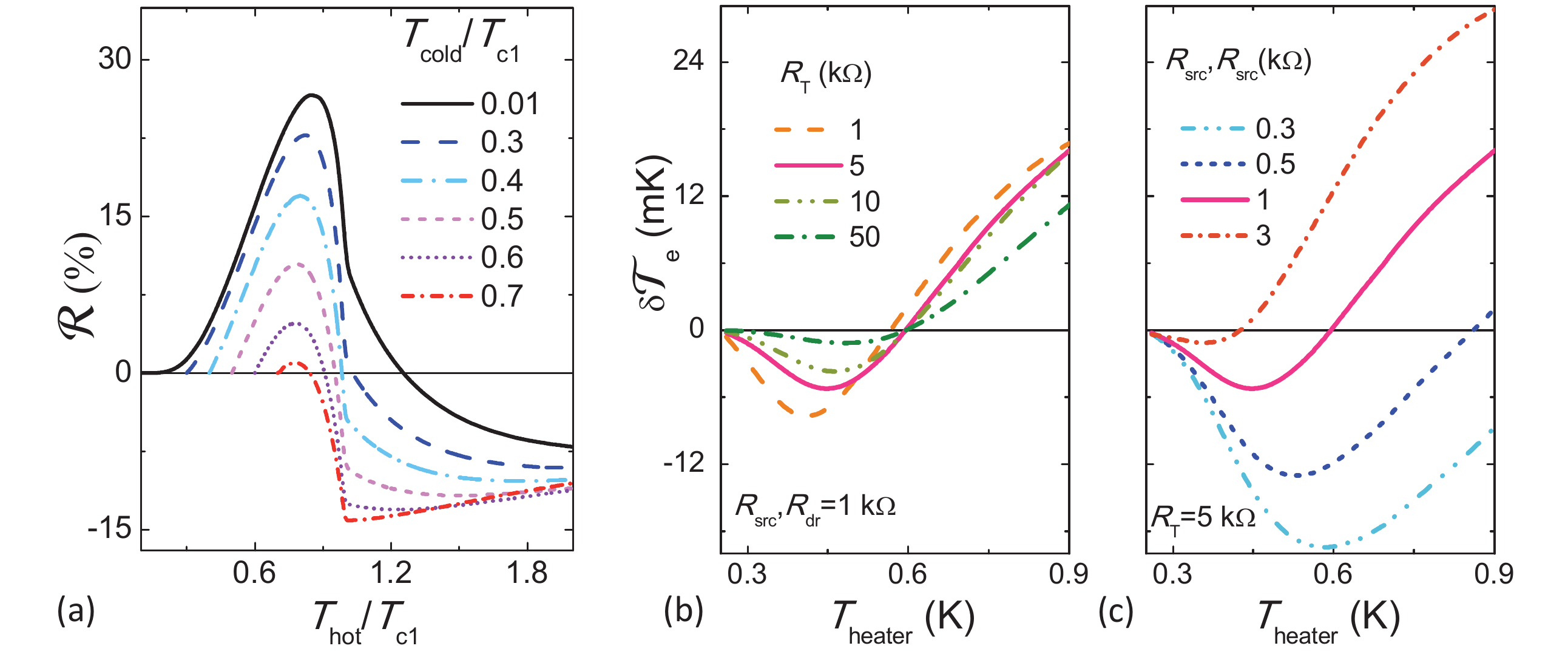}
\caption{(Color online) $\mathcal{R}$ vs. $T_{\rm hot}$ for different values of $T_{\rm cold}$ corresponding to a NIS thermal diode. Black horizontal line indicates $\mathcal{R}=0$.  (b) and (c) show the results from the thermal model described in Fig. \ref{Fig41}(d) corresponding to different fabrication parameters detailed in the legends. Black horizontal line indicates $\mathcal{T}_e=0$.}
\label{Fig42}
\end{figure}

  We calculate $\mathcal{R}$ as a function of $T_{\rm hot}$ for different values of $T_{\rm cold}$, i.e., for different temperature gradients established across the weak link. As shown in Fig.  \ref{Fig42}(c) a maximum positive rectification of $\sim 26 \% $ is obtained for $T_{\rm hot} $ aproaching $T_{c} $. %The latter means that the forward heat current is $\sim 1.3$ times larger than the reverse one. 
  As $T_{\rm hot}$ increases, heat rectification starts to decrease eventually inverting its sign, which implies that heat flux from N to S becomes preferred. Furthermore, by increasing $T_{\rm cold}$ leads to a reduction of $\mathcal{R}$, which reaches its maximum for larger values of $T_{\rm hot}$. Note that the heat rectification character of the NIS diode stems from the presence of two different DOS in the materials constituting the junction and the fact that the superconducting DOS is strongly temperature-dependent. As a consequence, $\mathcal{R}$ depends strongly on temperature as well, inverting its sign in the vicinity of $T_{\rm c}$, i.e., when the superconducting DOS becomes energy independent.  %Indeed, heat rectification vanishes when both junction electrodes are normal metals.  %A bare NIS junction offers, in this way, an already quite rich response in terms of heat rectification.

%____________________________________________________________________________________________________________________________________________________________
%____________________________________________________________________________________________________________________________________________________________
%____________________________________________________________________________________________________________________________________________________________
%____________________________________________________________________________________________________________________________________________________________

We conclude by analyzing the theoretical behavior of a realistic NIS-based thermal rectifier. For this purpose we consider the experimental design proposed in Fig. \ref{Fig41}(c). Two identical normal metal electrodes, source and drain, are weakly connected one each via resistances $R_{\rm src}=R_{\rm dr}$ to both S and N, respectively. Superconducting probes can be tunnel-coupled to these electrodes so to implement SINIS thermometers and heaters \cite{RMP}. Yet, the forward thermal bias configuration can be realized by intentionally increasing the electronic temperature in source electrode up to $T_{\rm src}^+ = T_{\rm heater} $ and probing the temperature in drain electrode $T_{\rm dr}^+$. On the reverse configuration, we set $T_{\rm dr}^- = T_{\rm heater} $ and  $T_{\rm src}^-$ is measured in a similar way. The difference $\delta \mathcal{T}_e= T_{\rm dr}^+ - T_{\rm src}^-$ for a given $T_{\rm h}$ can be used to assess  experimentally heat rectification. 

As we have done in the previous sections, $\delta \mathcal{T}_e $ can be computed numerically using the thermal model described in Fig. \ref{Fig41}(d). Thermal balance equations  %On the forward configuration, electrons in S exchange heat with electrons in the source at power $\dot{Q}_{\rm src}$ and, at power $\dot{Q}_{+}$,  with electrons in  N. On the other hand, electrons in N exchange heat with electrons in the drain at power $\dot{Q}_{\rm dr}$. Finally, electrons in the whole structure exchange heat at power $\dot{Q}_{\rm e-ph,\textit{k}}$ with lattice phonons that we assume to reside at bath temperature $T_{\rm bath}$, where $k=$src, S, N and dr. Under such circumstances, the three unknown quantities, i.e., $T_{\rm hot}$, $T_{\rm cold}$ and $T_{\rm dr}^+$ can be calculated for given initial conditions by solving the following system of thermal balance equations \cite{RMP} 
are given by $ \dot{Q}_{\rm src} (T_{\rm heater},T_{\rm hot})- \dot{Q}_{\rm e-ph,S}(T_{\rm hot}) -\dot{Q}_{+}(T_{\rm hot},T_{\rm cold})= 0$, $ \dot{Q}_{+}(T_{\rm hot},T_{\rm cold}) - \dot{Q}_{\rm e-ph,N}(T_{\rm cold})- \dot{Q}_{\rm dr}(T_{\rm cold},T_{\rm dr}^+)= 0$ and $ \dot{Q}_{\rm dr}(T_{\rm cold},T_{\rm dr}^+)-\dot{Q}_{\rm e-ph,dr}(T_{\rm bath},T_{\rm dr}^+)= 0$ for the $S$, $N$ and drain electrode, respectively\footnote{%In these expressions, $\dot{Q}_{\rm e-ph,\textit{l}} = \Sigma_{l} \mathcal{V}_{l} (T_{l}^{5}-T_{\rm bath}^{5})$ \cite{RMP} with $l=$src, N and dr, $\mathcal{V}_{l} $ and $\Sigma_{l} $ being the volume of the normal metal electrodes and the electron-phonon coupling constant, respectively. $\dot{Q}_{\rm e-ph,S}$, on the other hand, is that given in Eq.    (\ref{eph}). $\dot{Q}_{\rm src}$ is that given Eq. (\ref{qqp}) setting $\dot{Q}_{\rm src}=\dot{Q}_{qp}(T_{\rm heater}, T_{\rm hot})$ with $\mathcal{N}_2=1$ and substituting $R_{\rm J}$ for $R_{\rm src}$ and, finally, 
$\dot{Q}_{\rm dr}=\frac{2}{e^2 R_{\rm dr}}\int^{\infty}_{0} d\varepsilon \varepsilon [f(\varepsilon,T_{\rm dr}^+)-f(\varepsilon,T_{\rm cold})]$ \cite{RMP}. $T_{\textrm{src}}^-$ can be obtained in a similar way for the reverse configuration by simply exchanging the roles of $\dot{Q}_{\rm src} \Leftrightarrow \dot{Q}_{\rm dr} $, $\dot{Q}_{\rm e-ph,S} \Leftrightarrow \dot{Q}_{\rm e-ph,N}$, $\dot{Q}_{+} \Leftrightarrow \dot{Q}_{-}$ and $\dot{Q}_{\rm e-ph,dr} \Leftrightarrow \dot{Q}_{\rm e-ph,src}$. As representative  parameters we set $\mathcal{V}_{\rm src}=\mathcal{V}_{\rm dr}=\mathcal{V}_{\rm N}=\mathcal{V}_{\rm S} =  10^{-20}$ m$^{3}$. We choose Cu for which  $\Sigma_{\rm src}=\Sigma_{\rm dr}=\Sigma_{\rm N}\simeq 3 \times 10^9$ WK$^{-5}$m$^{-3}$ \cite{RMP} and Al $\Sigma_{\rm S} \simeq 0.3 \times 10^9$ WK$^{-5}$m$^{-3}$ \cite{RMP}. }. Using realistic parameters and assuming $T_{\rm bath}=245$ mK, the computed values of  $\delta \mathcal{T}_e $ vs. $T_{\rm heater}$  are plotted in Fig. \ref{Fig42}(b)  for different $R_{\rm T}$  values.  Remarkably, temperature differences exceeding $\sim 15$ mK can be obtained. As it can be seen,   $R_{\rm T}$  does not affect much the maximum value of $\delta \mathcal{T}_e $. The choice of $R_{\rm src}=R_{\rm dr}$, on the other hand, does influence noticeably the appearance and sign of  $\delta \mathcal{T}_e $ as shown in Fig. \ref{Fig42}(c).  The expected temperature differences  are easily measurable with standard Al-based SINIS or SNS thermometry techniques \cite{Giazottoarxiv,RMP}.

\subsection{SIS'-Josephson thermal diode}
\label{subsectionSIS}

%____________________________________________________________________________________________________________________________________________________________
%____________________________________________________________________________________________________________________________________________________________
\begin{figure}[t]
\centering
\includegraphics[width=0.65\columnwidth]{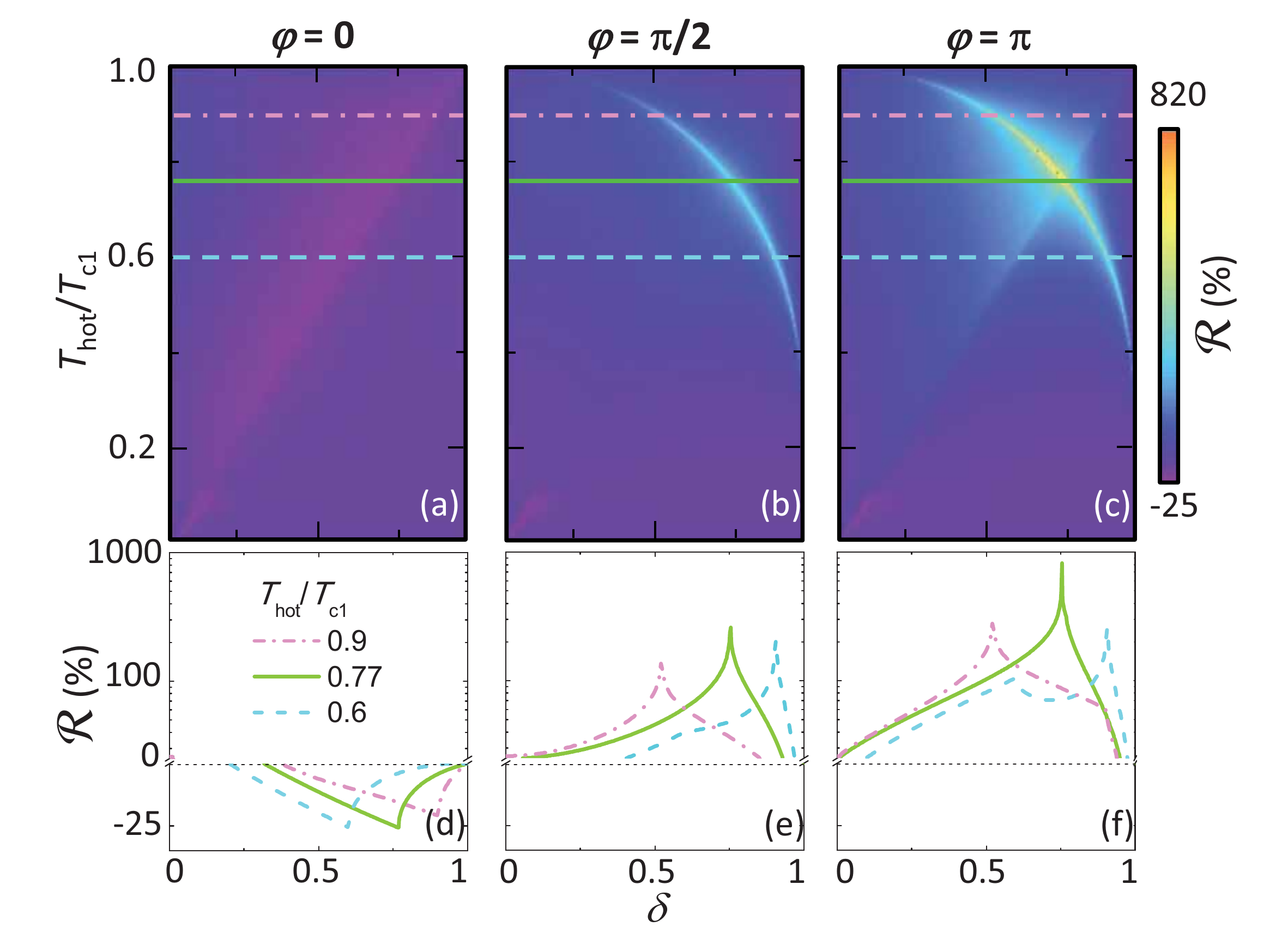}
\caption{(Color online) Panels (a), (b) and (c) show three density plots of $\mathcal{R}$  vs. $T_{\rm hot}$ and $\delta$ calculated for $\varphi = 0$, $\varphi = \pi / 2$ and $\varphi = \pi$, respectively.  Panels (d), (e) and (f) show three selected profiles of $\mathcal{R}$ vs $\delta$ for the same values of $\varphi$ corresponding to the colored straight lines in (a), (b) and (c). Notice that the scale is logarithmic above the break in the vertical axis. In addition, a dashed line indicates $\mathcal{R}=0$.   All curves have been calculated for $T_{\rm cold} = 0.01 T_{c_1}$. }
\label{Fig43}
\end{figure}
%____________________________________________________________________________________________________________________________________________________________
%____________________________________________________________________________________________________________________________________________________________

We consider finally the case of a SIS' junction for which we define $\delta= \Delta_1/ \Delta_2 \leq 1$ 
In this case, the forward and reverse total heat currents flowing through the JJ read \cite{GiazottoAPL12}
 $\dot{Q}_{+}= \dot{Q}_{qp}( T_{\rm hot}, T_{\rm cold}) -\dot{Q}_{int}(T_{\rm hot}, T_{\rm cold})\cos\varphi$ and 
 $\dot{Q}_{-}= -\Big[ \dot{Q}_{qp}( T_{\rm cold}, T_{\rm hot}) -\dot{Q}_{int}(T_{\rm cold}, T_{\rm hot})\cos\varphi\Big]$,
%\begin{equation}
%\begin{array}{lcll}
% \dot{Q}_{+}= \dot{Q}_{qp}( T_{\rm hot}, T_{\rm cold}) -\dot{Q}_{int}(T_{\rm hot}, T_{\rm cold})\cos\varphi, \\  \\
% \dot{Q}_{-}= -\Big[ \dot{Q}_{qp}( T_{\rm cold}, T_{\rm hot}) -\dot{Q}_{int}(T_{\rm cold}, T_{\rm hot})\cos\varphi\Big],
%\end{array}
% \label{forward} 
%\end{equation}
where the negative sign before brackets is set so that $\dot{Q}_{-}$ is positive by definition.

By fixing the temperature of the second electrode to $T_{\rm cold} = 0.01 T_{c_1}$, we calculate  $\mathcal{R}$ as a function of $T_{\rm hot}$ and as a function of $\delta$. The result is plotted in Fig. \ref{Fig43}(a), (b) and (c) for three representative cases, corresponding to $\varphi = 0$, $\varphi = \pi / 2$ and $\varphi = \pi$, respectively. Three selected profiles of $\mathcal{R}$ vs. $\delta$ for different $T_{\rm hot}$ values, i.e., for different thermal gradients, are shown as well in Fig. \ref{Fig43}(d), (e) and (f).  The inspection of these graphs reveals, on the first place, how phase biasing across the junction does make a substantial  difference. In particular, the heat rectification coefficient does not only change by almost two orders of magnitude from $\varphi = 0$ to $\varphi = \pi$ but it also switches its sign. It is worthwhile to emphasize that the SIS' junction rectifies heat only if $ \Delta_1 \neq \Delta_2 $. As for the case of the NIS diode, heat rectification demands the combination of two different DOS  being (at least one of them) strongly temperature-dependent. 

 Let us analyze in more detail the case for which  $\varphi = \pi$. In the forward configuration $\dot{Q}_{+}$ exhibits a local maximum when the temperature-dependent superconducting gaps of both superconducting electrodes coincide, which corresponds to the bright bend curve clearly visible in  Fig. \ref{Fig43}(c). On the other hand, for the reverse configuration, $\dot{Q}_{-}$ exhibit a local minimum when the $S_2$ electrode reaches its critical temperature, which corresponds to the straight line defined by $T_{1}/T_{c_1} = \delta$. As a result, $\mathcal{R}$ is maximized for $\Delta_1 (T_{\rm hot}) = \Delta_2 (T_{\rm cold})$ and $T_{1}/T_{c_1} = \delta$. In particular, $\mathcal{R} \sim 800 \%$ is reached at $\delta\simeq 0.75$ and $T_{\rm hot}\simeq0.77 T_{c_1}$\footnote{ For these calculations we have assumed a lifetime broadening $\gamma=10^{-5}$, which accounts well for the subgap leackage in realistic SIS junctions. $\mathcal{R}$ is reduced to $\sim 650 \%$ when considering $\gamma=10^{-4}$ and down to $\sim 500 \%$ for $\gamma=10^{-3}$.}.

\begin{figure}[t]
\centering
\includegraphics[width=0.9\columnwidth]{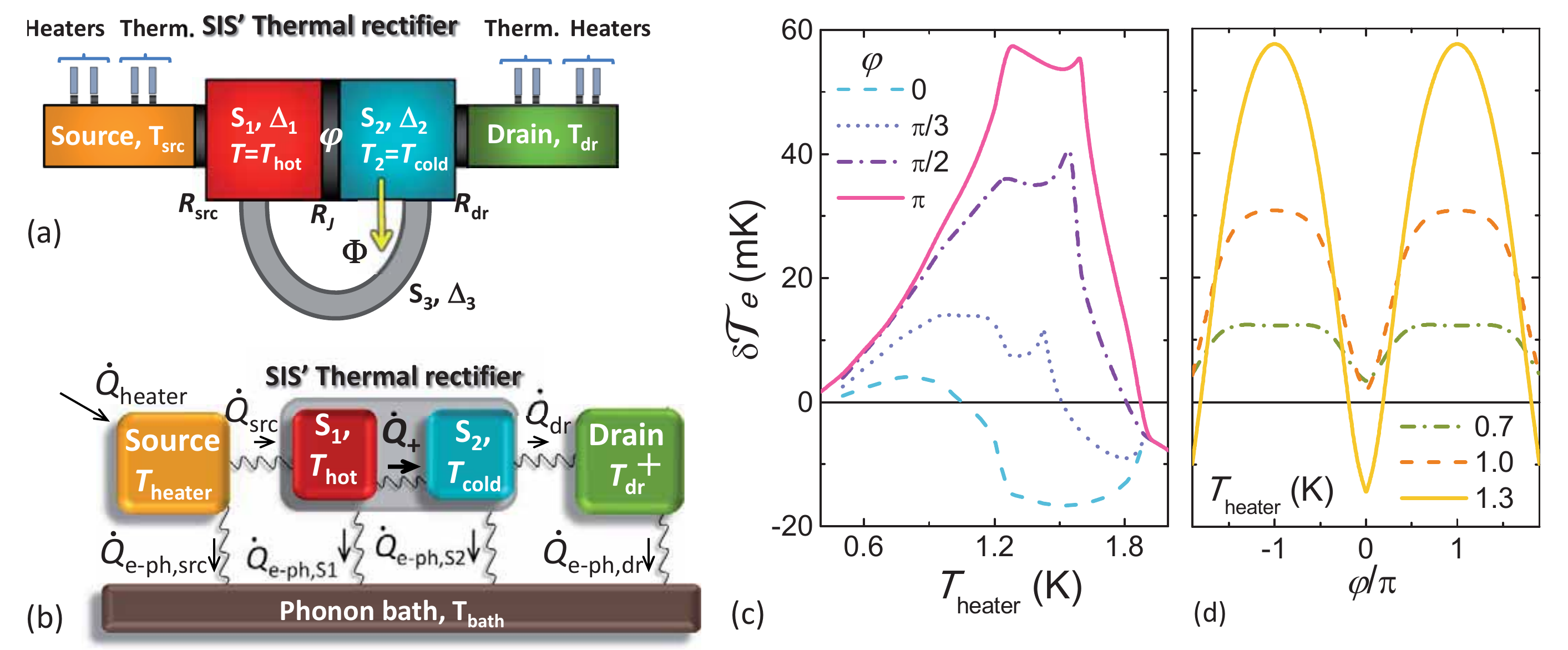}
\caption{(Color online) (a) Experimental realization of the SIS' thermal rectifier. Source and drain normal metal electrodes are tunnel-coupled to the core of the diode. Additional superconducting probes tunnel-coupled to source and drain allow for the implementation of SINIS thermometers and heaters. (c) Idealized thermal model including the main heat exchange mechanisms existing in the structure schematized in (a). Panels (c) and (d) show the computed values of $\delta \mathcal{T}_e $ as a function of $T_{\rm heater}$ and $\varphi$, respectively. Black horizontal line indicates $\delta \mathcal{T}_e $.}
\label{Fig45}
\end{figure}

The experimental realization of the SIS'-based thermal rectifier is similar to that of the NIS diode. In this case, we substitute N by a second superconducting electrode S$_2$ made, for instance, of Mn-doped Al since the latter allows for fine tuning of the aluminum superconducting gap \cite{Oneil}.  Optimum phase biasing can be achieved by using a rf SQUID configuration pierced by a control flux $\Phi$ [see Fig. \ref{Fig45}(a)]. 
For such a purpose, the thermal diode can be enclosed through clean contacts within a superconducting ring S$_3$ with energy gap $\Delta_3 \gg \Delta_1,\Delta_2$ so to suppress heat losses. Neglecting the inductance of the loop, the  phase-flux relation is given by $\varphi = 2 \pi \Phi/\Phi_0 $ \cite{Tinkham} enabling  the phase drop across the junction to vary within the whole phase space, i.e., $ -\pi  \leq \varphi \leq \pi$.  The experimental operation procedure will be that described in the previous section\footnote{ The thermal model is very similar to that used in the previous section but changing $\dot{Q}_{\rm e-ph,N}$ by $\dot{Q}_{\rm e-ph,S_2}$ as given in Eq. \ref{eph} and setting $\dot{Q}_{\rm dr}=-\dot{Q}_{qp}(T_{\rm cold}, T_{\rm dr}^{+})$ [Eq. (\ref{qqp})] with $\mathcal{N}_2=1$ and substituting $R_{\rm J}$ for $R_{\rm dr}$ [see Fig. \ref{Fig45}(b)]. As representative  parameters we set again $\mathcal{V}_{\rm src}=\mathcal{V}_{\rm dr}=\mathcal{V}_{\rm S_1}=\mathcal{V}_{\rm S_2} =  10^{-20}$ m$^{3}$, $R_{\rm J} = 5$ k$\Omega$ and $R_{\rm src} =R_{\rm dr}= 1$ k$\Omega$. We use Cu for which  $\Sigma_{\rm src}=\Sigma_{\rm dr}\simeq 3 \times 10^9$ WK$^{-5}$m$^{-3}$ \cite{RMP} and Al and Mn-doped Al with $\Sigma_{\rm S} \simeq 0.3 \times 10^9$ WK$^{-5}$m$^{-3}$ \cite{RMP}. For these materials we set $\Delta_2=1.4$ K and $\delta=0.75$.}. Solving the thermal model for $T_{\rm bath} = 245$ mK, the latter would give temperature difference approaching $\delta \mathcal{T}_e  \sim 60$ mK as it can be seen in Fig. \ref{Fig45}(c). Even more interesting, phase-coherence fingerprints are clearly observable as well. Notably, $\delta \mathcal{T}_e$ shows the expected $2\pi$-periodicity as shown in Fig. \ref{Fig45}(d).  These temperature differences  are easily measurable using  SNS thermometry techniques \cite{RMP} based on, for instance, vanadium with $T_c \approx 5$ K. 

%____________________________________________________________________________________________________________________________________________________________
%____________________________________________________________________________________________________________________________________________________________
%____________________________________________________________________________________________________________________________________________________________
%____________________________________________________________________________________________________________________________________________________________

%\section{Conclusions}
%\label{conclusions4}

\section{Summary and final remarks}
\label{Summary}

Along this manuscript we have proposed the experimental realization of different Josephson-based devices that exploit phase-coherence of heat currents to provide with a robust magnetic flux-to-temperature control. The successful realization of these architectures relies on the recent experimental demonstration of heat current interference in a simpler device. The latter consist of a ``thermal'' version of a conventional electric SQUID that provided with an experimental temperature modulation of $\sim 21$ mK in amplitude upon the application of an external magnetic field as reviewed in Sec. \ref{sec1}. 

In Sec. \ref{sec2} we have theoretically demonstrated how this modulation can be tuned \emph{in-situ} thanks to an improved design. Considering realistic parameters, such tunability leads to a much more robust drain temperature oscillations of $\sim 70$ mK. This is possible by replacing one of the SQUID JJ by an additional DC SQUID leading to a double-loop thermal interferometer that allows to maximize/minimize the strength of the phase-dependent component of the heat current. %The characteristics of this second ``junction'' can be tuned thanks to the application of an extra control magnetic flux that behaves as a second control knob for the magnetic flux-to-temperature transfer function. 
The existence of two control knobs, i.e., two externally applied magnetic fluxes, may be exploited to perform non-trivial adiabatic cycles in its control space eventually leading to the realization of a heat pump \cite{ren10}.

In Sec. \ref{sec4} we have shown that the heat current through a temperature-biased extended JJ under the influence of an  in-plane magnetic field displays \emph{coherent diffraction}, in full analogy with the Josephson \emph{electric} critical current \cite{rowell}. %We have proposed and analyzed a hybrid superconducting microstructure, easily implementable with current technology, which would allow to demonstrate diffraction of thermal currents. 
Depending on the junction geometry, the latter will lead to a plethora of magnetic flux-to-temperature diffraction patterns with modulation amplitudes in the range of $30 - 60$ mK. The experimental realization of this device has been reported very recently \cite{MartinezNat} setting a complementary and conclusive demonstration of the ``thermal'' Josephson effect in weakly-coupled superconductors. % as it was done 50 years ago for its ``electric'' counterpart. Here, we refer to the experiments made by J. M. Rowell in 1963 in which the single-slit diffraction pattern of the critical current flowing through a JJ was measured for the first time \cite{rowell} therefore providing strong experimental evidence of the DC \emph{electric} Josephson effect \cite{Josephson}.

We have finally proposed and analyzed in Sec. \ref{sec5} the concept of a NIS-based and a SIS' Josephson thermal rectifier. Under appropriate conditions, remarkably large rectification coefficients of $\mathcal{R} \sim 30 \%$ and $\mathcal{R} \sim 800 \%$ can be obtained for the NIS and SIS' diodes, respectively. Combining the diode within a realistic device, temperture differences between the forward and reverse configurations of the order of $15$ and $60$ mK could be reached for the NIS and SIS' diodes, respectively.  In addition, thanks to phase-coherence, the rectification character of the SIS' diode can be maximized or even inverted \emph{ in-situ}. Such a device might find a straightforward application of technological interest, e.g., in the field of electronic refrigeration enabling magnetic-flux dependent heat management and thermal isolation at the nanoscale. %The operation principle which is at the basis of this heat rectifier will likely contribute, on the other hand, to improve the performance of other different coherent thermal components such as the double-loop or the Josephson quantum diffractor analyzed in Sec. \ref{sec2} and \ref{sec4}, respectively.

The structures described here can be integrated within, not only superconducting elements, but also hybrid mesosocopic circuits composed of, e.g., normal metals, two dimensional electron gases and semiconductor nanowires as well. In this way, many fields of research such as radiation detectors or quantum computing  might benefit from our approach \cite{RMP}. %In particular, two disciplines for which the mastering of heat would represent an important breakthrough are the field of radiation detectors and the field of quantum computing based on mesoscopic superconducting circuits. %The electronic temperature in superconductors determines ultimately its phase transition to the normal state \cite{Tinkham}. This property has led to the development of transition edge sensors TES for radiation detection \cite{RMP}. The devices proposed here could enable for a faster in-situ fine tuning of the electronic temperature in such detectors by simply applying an external magnetic field tuned in a microsecond timescale by using fast on-chip flux lines. Superconducting quantum computing architectures could also benefit enormously of these technology as the decoherence time of quantum bits (qubits) is directly related to temperature fluctuations \cite{NielsenChuang}. The electronic temperature determines, on the other hand, the energy level occupation in quantum systems. Controlling the temperature of superconducting qubits could eventually contribute to its selective initialization. 
%Finally, the versatility of general-purpose superconducting microcircuits can also benefit from these fully tunable Josephson-based thermal circuits. In such devices, the direct relation between the electronic temperature and the critical current can be exploited to modulate the latter via the application of a magnetic flux. Unlike the usual voltage-controlled hot-electron Josephson transistors \cite{Morpurgo}, the devices proposed here can lead to magnetic fux-controlled thermal Josephson transistors.
On the other hand, these devices represent the first step towards the realization of coherent caloritronic circuits. The discipline usually referred to as \emph{caloritronics} (from \emph{calor-}, i.e., ``heat'' in latin + \emph{electronics}), deals with the generation and mastering of heat currents through an electronic drive. %Yet, in superconducting circuits based on Josephson junctions, heat can be additionally manipulated in a coherent fashion. %Hence, the technology foreseen here represents the first step towards the birth of coherent caloritronic circuits. Just as electronic circuits consist of a number of components (e.g., transistors, diodes and switches) connected together to enable the execution of different operations, coherent caloritronic circuits might eventually allow for the implementation of, for instance, thermal computation and thermal logic operations. 
Yet, our approach provides with the possibility of manipulating heat currents to flow through a series of superconducting components exploiting their coherent character.

The core papers described in this review deal only with static parameter configurations where the superconducting phase and temperature are time-independent.
%Their interest lies in the confirmation of a 50-year-old prediction that a fraction of the heat flux between superconductors depends on the superconducting phase difference and, therefore, it can be controlled through magnetic fluxes. Once established this connection, 
The natural extension is to control and change the superconducting phase in time  by applying a magnetic field. 
In principle this opens the possibility to transfer dynamically the heat between the superconductors, cool or heat one them and build a thermal mesoscopic engine.
This new field requires both theoretical and experimental advance and will be the subject of future research.

\begin{acknowledgements}
The FP7 program No. 228464 MICROKELVIN, the Italian Ministry of Defense through the PNRM project TERASUPER, the Marie Curie Initial Training Action (ITN) Q-NET 264034 and FIRB - Futuro in Ricerca 2012 under Grant No. RBFR1236VV HybridNanoDev. are acknowledged for partial financial support
\end{acknowledgements}

%\bibliographystyle{spr-chicago}      % Chicago style, author-year citations
%\bibliography{example}   % name your BibTeX data base
%\nocite{*}


\end{document}